%% file: phep-2012-350-arxiv.tex
\documentclass[manyauthors,nocleardouble,COMPASS]{cernphprep}
\usepackage{graphicx}
\usepackage{subfig}
\usepackage[percent]{overpic}
\usepackage[numbers, square, comma, sort&compress]{natbib}

\usepackage{ifpdf}

\usepackage[english]{babel}

\usepackage{graphicx}
\usepackage{rotating}
\usepackage{multirow}
\usepackage{amsmath}
\usepackage{array}

\usepackage{hyperref}

\newcommand{\DG }{$\left< {\Delta g}/{g} \right>$}
\newcommand{\Dg }{$\left< {\Delta g}/{g} \right>$~}

\newcommand{\GeV}{ \mathrm{GeV}}

\def\lapproxeq{\lower .7ex\hbox{$\;\stackrel{\textstyle
<}{\sim}\;$}}
\def\gapproxeq{\lower .7ex\hbox{$\;\stackrel{\textstyle
>}{\sim}\;$}}

\begin {document}
\selectlanguage{english}

\begin{titlepage}
\PHnumber{2012--350}
\PHdate{November 26, 2012}

\title{Leading and Next-to-Leading Order Gluon Polarisation 
in the Nucleon and Longitudinal Double Spin
Asymmetries from Open Charm Muoproduction}
  
\Collaboration{The COMPASS Collaboration}
\ShortAuthor{The COMPASS Collaboration}

\begin{abstract}
The gluon polarisation in the nucleon was measured using open charm
production by scattering 160 GeV/$c$ polarised muons off
longitudinally polarised protons or deuterons. The data were taken by
the COMPASS collaboration between 2002 and 2007. A detailed account is
given of the analysis method that includes the application of neural
networks.  Several decay channels of D$^0$ mesons are investigated.
Longitudinal spin asymmetries of the D meson production cross-sections
are extracted in bins of D$^0$ transverse momentum and energy. At
leading order QCD accuracy the average gluon polarisation is
determined as \DG$^{\rm LO}=-0.06 \pm 0.21 ~(\mbox{stat.}) \pm 0.08
~(\mbox{syst.})$ at the scale $\langle\mu^2\rangle \approx
13\,(\GeV/c)^2$ and an average gluon momentum fraction $\langle
x_{\rm} \rangle \approx 0.11$.  For the first time, the average gluon
polarisation is also obtained at next-to-leading order QCD accuracy as
\DG$^{\rm NLO}=-0.13 \pm 0.15 ~(\mbox{stat.}) \pm 0.15
~(\mbox{syst.})$ at the scale $\langle\mu^2\rangle \approx
13\,(\GeV/c)^2$ and $\langle x_{\rm} \rangle \approx 0.20$.
\end{abstract}

\vfill
\Submitted{(submitted to The Physical Review D)}
\end{titlepage}

{\pagestyle{empty}
\input{Authors2012CharmDG.tex}
\clearpage
}

\maketitle

\section{Introduction}
\label{sec:intro}
The decomposition of the nucleon spin projection of 1/2 (in units of
$\hbar$) into contributions from helicities and orbital angular
momenta of partons became a topic of major interest in experimental
and theoretical hadron physics after the European Muon Collaboration
at CERN had published the surprising result that quark helicities
contribute only an unexpectedly small fraction~\cite{emc}.  Extensive
nucleon spin studies were carried out at CERN \cite{smc,compass}, SLAC
\cite{e155_d}, DESY \cite{hermes}, JLAB \cite{jlab} and at BNL
\cite{phenix,star}.  From the data, parton helicity distributions of
the nucleon were extracted using the framework of perturbative QCD. By
now, the contribution of the quark helicities to the nucleon spin is
known to be about 30\%, significantly smaller than the value of 60\%
expected from the Ellis--Jaffe sum rule~\cite{refa8}. Relativistic
quark motion is responsible for the reduction from the value of 100\%,
expected in the na\"ive quark-parton model \cite{bass}. In spite of
the ongoing theoretical debate on how to correctly perform a
gauge-invariant decomposition of the nucleon spin, agreement exists
that besides the contributions of the quark helicities also the gluon
helicity contribution $\Delta G$ is a measurable, gauge-invariant
observable (see, {\it e.g.} Ref. \cite{leader}).

The data of present inclusive polarised deep-inelastic scattering
experiments cover a narrower range in the photon virtuality, $Q^2$, as
compared to unpolarised ones and hence their QCD analyses ({\it e.g.}
Ref. \cite{lss_2010}) show only limited sensitivity to the gluon
helicity distribution as a function of the gluon momentum fraction
$x$~\footnote{Throughout this paper, $x$ denotes the gluon momentum
fraction, while $x_{\rm B}$ stands for the Bjorken scaling variable.},
$\Delta g(x)$, and to its first moment, $\Delta G$.  Such a
determination of $\Delta g(x)$ from QCD evolution has therefore to be
complemented by direct, dedicated measurements.

Direct determinations of the average gluon polarisation in a limited
interval of $x$, \DG, were performed in a model-dependent way using
the Photon-Gluon Fusion (PGF) process by SMC \cite{SMC_highpt}, HERMES
\cite{hermes_highpt} and COMPASS \cite{compass_highpt_lowq}. These
analyses used events containing hadrons or hadron pairs with high
transverse momenta, typically 1 to 2~GeV/$c$.  This method provides
good statistical precision but relies on Monte Carlo generators
simulating QCD processes.  PYTHIA \cite{pythia} was used by HERMES and
by COMPASS for the analysis of small $Q^2$ events, and LEPTO
\cite{lepto} by SMC and COMPASS for $Q^2>1~(\GeV/c)^2$ events
\cite{marcin}.  All measurements yield a small value of the gluon
polarisation at $x\approx 0.1$. This is consistent with recent results
from PHENIX \cite{phenix} and STAR \cite{star} at RHIC, where the
production of inclusive $\pi^0$ or high transverse momentum jets led
to constrain the magnitude of \DG.

\begin{figure}[!h]
\begin{center}
\vskip-3cm\includegraphics[width=0.6\textwidth]{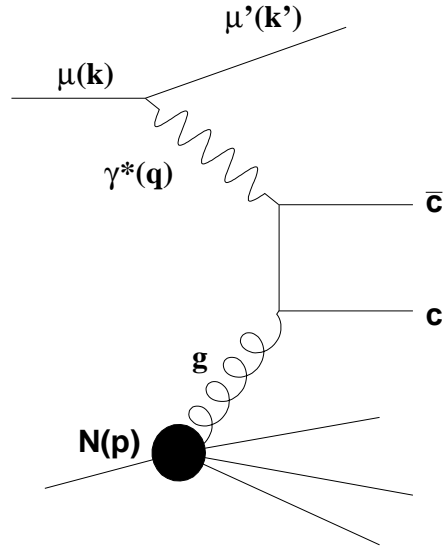}
\vskip-1cm\caption{Photon-Gluon Fusion into a pair of charm quarks,
$c\bar c$. Symbols in parentheses denote four-vectors.}
\label{fig:pgf_diagram}
\end{center}
\end{figure}

In this paper, we present new results on \Dg and the virtual
photon-nucleon asymmetries obtained from charm production tagged by D
meson decays in 160 GeV/$c$ polarised muon-nucleon scattering.  The
data were collected by the COMPASS Collaboration at CERN in the
2002--2004 and 2006--2007 running periods.  The results supersede the
ones given in Ref. \cite{compass_pgf} since they are based on the full
data sample and an improved analysis method; additional final state
channels are added as well.  The gluon polarisation is determined
assuming that open charm production is dominated by the PGF mechanism,
$\gamma^*g \rightarrow c\bar{c}$, as depicted in
Fig.~\ref{fig:pgf_diagram}.  The subsequent fragmentation of the
$c\bar{c}$ pair, mainly into D mesons, is assumed to be
spin-independent.  The dominance of the PGF mechanism in the COMPASS
kinematic region is supported by the EMC results on $F_2^{\rm
c\bar{c}}$ (Ref. \cite{emccharm}, further discussed in
Ref. \cite{harris}), and by a COMPASS study of charm meson production
\cite{martin}.  The determination of the gluon polarisation based on
this assumption, although limited statistically, has the advantage
that in lowest order of the strong coupling constant there are no
other contributions to the cross section.

In the present analysis, only one charmed meson is required in every
event. This meson is selected through its decay in one of the
following channels: ${\rm D}^{*}(2010)^+ \rightarrow {\rm
D^0}\pi^+_{\rm slow}\rightarrow ({\rm K^-}\pi^+/{\rm
K^-}\pi^+\pi^0/{\rm K^-}\pi^+$ $\pi^+\pi^-)\pi^+_{\rm slow}$ or ${\rm
D^0}\rightarrow {\rm K^-}\pi^+$, as well as their charge
conjugates. The former samples are called `tagged' ones while the
latter is denoted `untagged'. Virtual photon cross section
asymmetries, $A^{\gamma {\rm N}\rightarrow {\rm D}^0{\rm X}}$, and the
average gluon polarisation \Dg are extracted from these open charm
events. In Table \ref{table:variables}, the kinematic variables
describing the $\mu$N scattering process are listed.  In this analysis
we have also employed, for the first time, next-to-leading order QCD
calculations for the determination of the gluon polarisation.  Since
the PGF process is dominated by quasi-real photoproduction ($Q^{2}
\rightarrow 0$), the perturbative scale for the selected events,
$\mu^2$, cannot be set to $Q^2$ as in the QCD analyses of inclusive
data. Instead, this scale is chosen to be the transverse mass of the
charmed quarks, $\mu^2 \equiv 4M^2_{\rm T} = 4(m^2_c + p^2_{\rm T})$,
where the D meson transverse momentum, $p_{\rm T}$, is defined with
respect to the virtual photon.

\begin{table}[!h]
\caption[Table caption]{Kinematic variables used to describe the muon-nucleon
interaction, see  Fig. \ref{fig:pgf_diagram}. 
\label{table:variables} }
\begin{center}
\begin{tabular}{|l|c|l|}
\hline
&&\\
~~~~~~~~~~~~~~~~~{\bf Variable}&{\bf Symbol}&{\bf ~~~~~~~~Definition}\\
&&\\
\hline
Nucleon (muon) mass&$M(m_\mu)$&\\
Four vector of incoming (outgoing) muon&$k(k^\prime)$&\\
Four vector of target nucleon&$p$&\\
Four vector of outgoing hadron final state&$p_{\rm X}$&\\
Four vector of virtual photon&$q$&$q = k - k^\prime$\\
Four vector of a final state hadron&$p_{\rm h}$&\\
&&\\
Negative four momentum transfer squared&$Q^2$&$Q^2=-q^2$\\
(photon virtuality)&&\\
Muon laboratory incident (final)  energy&$E(E^\prime)$&$E=\displaystyle\frac{p\cdot k}{M}
~~\left (E^\prime=\frac{p\cdot k^\prime}{M}\right )$\\
&&\\
Polar angle of kaon in the D$^0$ centre-of-mass &$\theta^*$&\\relative to the D$^0$ laboratory momentum&&\\
&&\\
Virtual photon energy&$\nu$&$\nu = \displaystyle\frac{p\cdot q}{M} 
\overset {\rm lab}{~=~} E - E^\prime$\\
&&\\
Bjorken scaling variable
&$x_{\rm B}$&$x_{\rm B} = 
\displaystyle\frac{Q^2}{2p\cdot q} \overset {}{~=~} \frac{Q^2}{2M\nu}$\\
&&\\
Virtual photon fractional energy 
& $y$&$y = \displaystyle\frac{p\cdot q}{p\cdot k}
\overset {\rm lab}{~=~} \frac{\nu}{E}$\\
&&\\
Final state hadron fractional energy& $z_{\rm h}$&$z_{\rm h} = 
\displaystyle\frac{p\cdot p_h}{p\cdot q} 
\overset {\rm lab}{~=~}\frac{E_h}{\nu}$\\
&&\\
Transverse  momentum of D$^0$ meson&  $p_{\rm T}^{{\rm D}^0}$&\\
with respect to the virtual photon direction&&\\
&&\\
Energy of  D$^0$ meson in laboratory&$E_{{\rm D}^0}$&\\ 
&&\\
\hline
\end{tabular}
\end{center}
\end{table}

The paper starts with a brief presentation of the experiment
in Sec. \ref{sec:experiment}. In Sec. \ref{sec:data_selection} 
the data selection is reported in detail. The
evaluation of the asymmetries and the corresponding results are
described in Sec. \ref{sec:Method}.
The determinations of the gluon polarisation $\langle\Delta g/g\rangle$ 
at leading (LO)  and next-to-leading (NLO) QCD accuracies 
are presented in Sec.~\ref{sec:results_tot}. 
Concluding remarks are given in Sec. \ref{sec:conclusions}.

\section{Experimental set-up}
\label{sec:experiment}
The COMPASS spectrometer is a fixed target set-up situated at the M2
beam line of the CERN SPS using muon or hadron beams.  For the present
measurement, longitudinally polarised positive muons of 160~GeV/$c$
momentum were scattered off a large polarised solid state target.  A
detailed description of the set-up can be found in
Ref. \cite{spectrometer}.

The muons originate from the weak decay of 175 GeV/$c$ pions and kaons
produced by the 400~GeV/$c$ SPS proton beam impinging on a primary
beryllium target and are thus naturally polarised. The beam
polarisation, $P_{\mu}$, is about 0.8 at 160~GeV/$c$ with a relative
uncertainty of 5\% \cite{beampol}. A beam intensity of about $4\cdot
10^7$~muons/s was used, with the spill length between 4.8~s and 9.6~s
for SPS cycles between 16.8~s and 48~s, respectively.  The beam is
focused onto the target centre with a spread of 7~mm (r.m.s.) and a
momentum spread of 5\% for the Gaussian core.  The momentum of each
incoming muon is measured with a precision better than 1\% upstream of
the experimental hall using a beam momentum station.  Before the
target, the trajectory of each beam particle is determined with an
angular precision of 30 $\mu$rad using a set of scintillating fibre
and silicon detectors.

The solid state target is housed in a large superconducting solenoid
providing a field of 2.5~T with field uniformity, $\delta B /B$,
better than $10^{-4}$.  From 2002 to 2004, the angular acceptance was
$\pm69$~mrad at the upstream edge and $\pm170$~mrad at the downstream
edge of the target material.  From 2006 onwards, a new target magnet
with a larger aperture solenoid was used \cite{spectrometer}. It
yields an angular acceptance of $\pm180$~mrad for the upstream target
edge resulting in a much improved hadron acceptance and matching the
$\pm180$~mrad acceptance of the spectrometer.

The target material consisted of $^6$LiD beads in 2002 to 2006 and
NH$_3$ beads in 2007, in a bath of \mbox{$^3$He -$^4$He}.  The target
was cooled down to a temperature below 100~mK by a $^3$He -$^4$He
dilution refrigerator. The target polarisation was accomplished using
the method of dynamic nuclear polarisation (DNP) and measured
continuously by a set of NMR coils surrounding the target material.
The achieved polarisation, $P_{t}$, was about 0.5 for deuterons
($^6$LiD) and 0.9 for protons (NH$_3$) with a relative uncertainty of
5\% and 2\%, respectively.

In 2002 to 2004, the target material was contained in two 60~cm long
cells that were polarised in opposite directions. The polarisation was
reversed 3 times per day by rotating the field of the target
magnet. From 2006 onwards, a three-cell target set-up was used with a
central 60~cm long cell placed between two 30~cm long ones.  The
material inside the central cell was polarised oppositely to that of
the outer ones. The use of this new target arrangement allows for
further reduction of the systematic uncertainty due to the variation
of the spectrometer acceptance along the target, so that only one
field rotation per day was performed.  In order to minimise possible
acceptance effects related to the orientation of the solenoid field,
also the sign of the polarisation in each target cell was reversed
several times per year by changing the DNP microwave frequencies.

As not all nucleons in the target material are polarised, the
so-called dilution factor, $f$, is introduced.  It is expressed in
terms of the numbers $n_{\rm A}$ of nuclei with mass number $A$ and
the corresponding total ({\it i.e.} including radiative effects)
spin-independent cross sections, $\sigma_{\rm A}^{\rm tot}$, per
nucleon for all the elements involved:
$$f_{\rm H,D} = \frac{n_{\rm H,D}\cdot \sigma_{\rm H,D}^{\rm tot}}
{\displaystyle\Sigma_{\rm A} n_{\rm A}\cdot \sigma_{\rm A}^{\rm
tot}}.$$ In the present analysis, the dilution factor is modified by a
correction factor $\rho= \sigma^{1\gamma}_{\rm p,d}/\sigma^{\rm
tot}_{\rm p,d}$ accounting for the dilution due to radiative events on
unpolarised protons (deuterons) \cite{radcorr}.  A correction for
polarisation of the deuteron in the $^6$Li nucleus is also applied.
 
The dilution factor depends on $x_{\rm B}$. At low $x_{\rm B}$, it is
larger for events containing hadrons in the final state due to the
absence of radiative elastic tails. Its values at medium $x_{\rm B}$
for $^6$LiD and NH$_3$ are about 0.37 and 0.14 with relative
uncertainties of 2\% and 1\%, respectively .

The two stages of the COMPASS set-up are open dipole spectrometers for
large and small angle tracks, respectively.  Each dipole magnet is
surrounded by tracking detectors. COMPASS uses various types of them
in order to match the expected particle flux at various locations in
the spectrometer. In high-flux regions close to the beam, tracking is
provided by arrays of scintillating fibers, silicon detectors,
micromesh gaseous chambers and gas electron multiplier
chambers. Further away from the beam, larger-area tracking devices as
multiwire proportional chambers, drift chambers and straw detectors
are used. In 2006 the tracking system in the first stage of the
spectrometer was adapted to match the increased aperture of the
superconducting solenoid.

Muons are identified in large area tracking detectors and
scintillators downstream of concrete or iron muon filters. Hadrons are
detected by two scintillator-iron sandwich calorimeters installed in
front of the muon filters.  Electromagnetic lead glass calorimeters
are placed in front of the hadron ones.  The data recording system is
activated by triggers indicating the presence of a scattered muon
and/or energy deposited by hadrons in the calorimeters. Both inclusive
and semi-inclusive triggers are used. In the former, the scattered
muon is identified by coincident signals in the trigger hodoscopes,
and in the latter the energy deposited in calorimeters is demanded in
addition.  Moreover, a calorimetric trigger with a high energy
threshold is implemented to extend the acceptance. In order to
suppress triggers due to halo muons, veto counters upstream of the
target are used.  The COMPASS trigger system covers a wide range in
$Q^2$, from quasi-real photoproduction to the deep inelastic region.

For charged particle identification in the first stage of the set-up,
a~Ring Imaging Cherenkov detector (RICH) is installed
\cite{rich_paper}.  It is a gas RICH with a 3~m long C$_4$F$_{10}$
radiator. Two spherical mirror surfaces reflect and focus the
Cherenkov photons on two sets of detectors situated above and below
the acceptance of the tracking detectors, respectively.  The photon
detection uses MWPCs with segmented CsI photocathodes which detect
photons in the UV region. In 2006, the central part of the RICH was
upgraded replacing the MWPCs by multianode photomultiplier tubes
yielding a considerably higher number of detected photons and a much
faster response. For the outer parts, the readout electronics was
refurbished allowing a significant reduction of the background.

The particle identification procedure relies on a likelihood function
based on information on the photons detected in the RICH and
associated with a charged particle trajectory.  The likelihood
function uses the photons of the signal and a theoretical expectation
of their distribution, taking into account possible signal losses due
to dead zones in the detector.  For the description of the background
photons, the experimental occupancy of the photon detectors is used.
For each track, likelihood values are computed for different particle
mass hypotheses and the background hypothesis.  Identification of a
pion (kaon) is possible for momenta between 2.5 GeV/$c$ (9 GeV/$c$)
and 50 GeV/$c$.

The performance of the detectors as well as the stability of the
reconstructed data was carefully monitored and all spills not
fulfilling stability requirements were excluded from further analysis.
Time intervals selected for asymmetry measurements correspond to
periods of stable spectrometer performance.  In total, data taking
amounted to 48 weeks in the years 2002 to 2007.

\section{Data selection}
\label{sec:data_selection}
In order to extract information about the gluon polarisation, events
with D mesons have to be selected from the data.  This is accomplished
by requiring every event containing an incoming and outgoing muon
together with at least two outgoing charged tracks.  Furthermore, only
events with the incoming muon potentially crossing the whole target
and with an interaction point (or `vertex') within the target were
retained.

\begin{figure}[!h]
\begin{center}
\includegraphics[totalheight=6.0cm,width=1.05\textwidth]{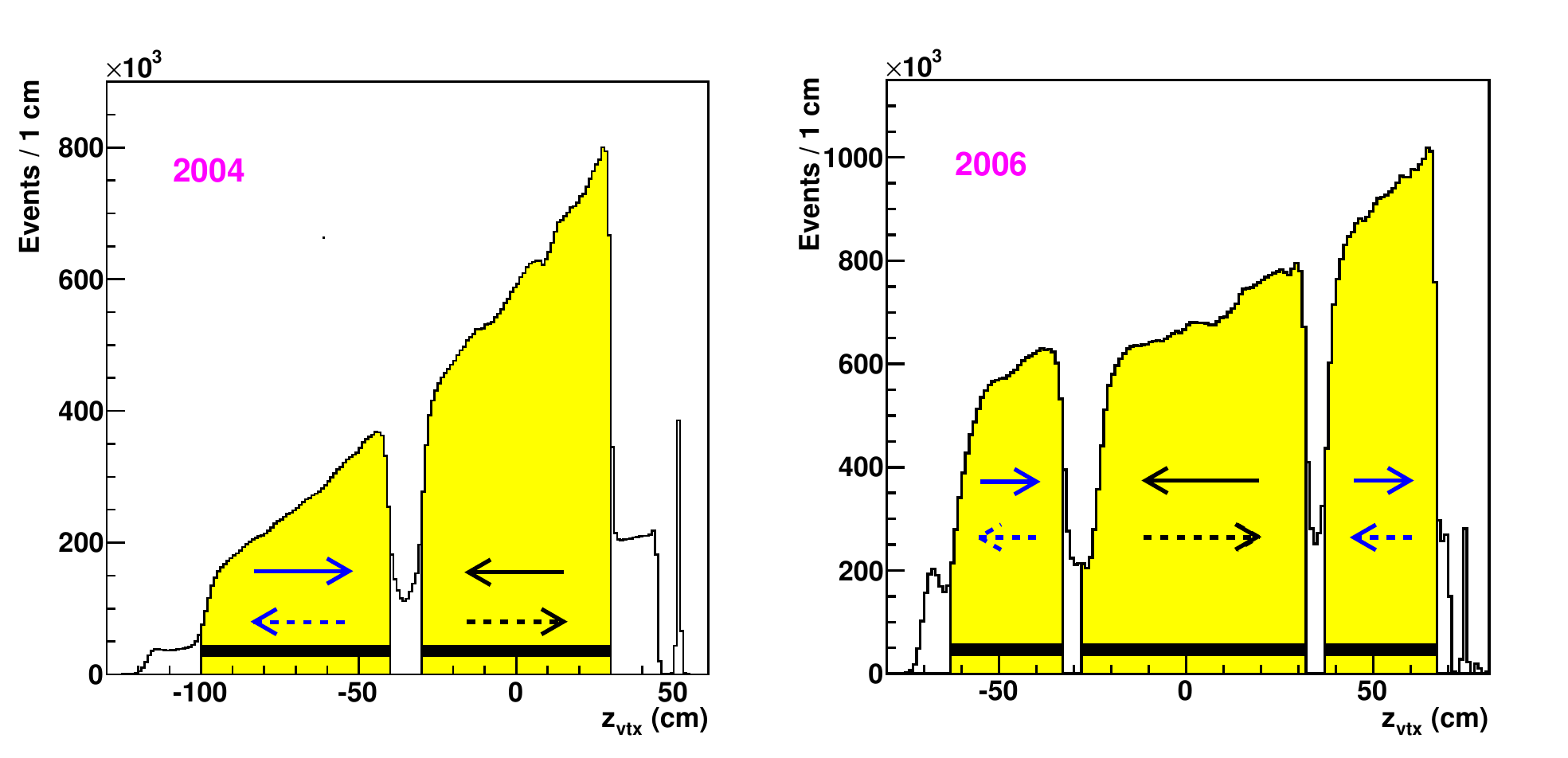}
\caption{Distribution of reconstructed vertex positions 
$z_{\rm vtx}$ along the
beam axis for the target with two (left) and three (right) cells. 
Dark horizontal bars at the bottom mark the target extensions, arrows 
denote the target polarisation directions. See text for details.}
\label{fig:z_distribution_for_2_3_cells}
\end{center}
\end{figure}

 The direction of tracks reconstructed at an interaction point in the
target is determined with a precision better than 0.2~mrad and the
momentum resolution for charged tracks detected in the first (second)
spectrometer stage is about 1.2\% (0.5\%). The longitudinal vertex
resolution varying from 5~mm to 25~mm along the target permits
assigning each event to a particular target cell, {\it i.e.} to a
specific target spin direction.  In
Fig. \ref{fig:z_distribution_for_2_3_cells} are shown the
distributions of the reconstructed vertex position $z_{\rm vtx}$ along
the beam axis for events remaining after applying the aforementioned
selection criteria.  The relative increase of the number of events in
the upstream and central target cells, seen in the right panel,
reflects the acceptance increase due to the upgrade of the target
magnet in 2006.

Due to multiple Coulomb scattering in the solid state target, the
spatial resolution of the vertex reconstruction is not sufficient to
separate production and decay points of charmed mesons. As a result,
such mesons can only be reconstructed using the invariant mass of
their decay products.  Their decay modes considered in this analysis
are listed in Table \ref{table:decay_modes}.  Those D$^0$ decays which
involve the same set of final state particles cannot be distinguished
event by event.  Therefore five independent data samples with
different final states are defined for this analysis, see Table
\ref{table:channels}.  In the four tagged samples, {\it i.e.}
D$^{*}_{{\rm K}\pi}$, D$^{*}_{{{\rm K}_{\rm sub}}\pi}$, D$^{*}_{{\rm
K}\pi\pi^{0}}$, and D$^{*}_{{\rm K}\pi\pi\pi}$, the D$^0$ meson is
assumed to originate from a D$^*$ decay into a D$^0$ meson and a slow
pion, ${\rm D}^{*} \
\overset{\underset{\mathrm{67.7\%}}{}}{\longrightarrow} \ {\rm
D}^{0}\pi_{\rm s}$.
The kinematic selection criteria, which are tuned to reduce the
combinatorial background without affecting the D$^0$ meson signal, are
listed in Table \ref{table:kin_cuts}.

\begin{table}[h]
\begin{center}
\caption[Table caption]{Charmed D$^0$ meson decay modes,
together with their branching ratios, considered in this analysis. 
The charge conjugate (c.c.) final states from  $\overline{{\rm D}^0}$
decays are also included.\label{table:decay_modes} }
\begin{tabular}{|c|lc|}
\hline
&&\\
{\bf Reaction}&\multicolumn{2}{|c|}{\bf D$^0$ decay mode}\\
{\bf number}&&\\
&&\\
\hline
1&${\rm D}^{0} \ \overset{\underset{\mathrm{3.89\%}}{}}{\longrightarrow} 
\ {\rm K}^{-}\pi^{+} \ $
&$+ \ {\rm c.c.}$
\\
2&${\rm D}^{0} \ \overset{\underset{\mathrm{13.9\%}}{}}{\longrightarrow} 
\ {\rm K}^{-}\pi^{+}\pi^{0} \ $ 
&$+ \ {\rm c.c.}$
\\
3&${\rm D}^{0} \ \overset{\underset{\mathrm{8.09\%}}{}}{\longrightarrow} 
\ {\rm K}^{-}\pi^{+}\pi^{+}\pi^{-} \ $
&$+ \ {\rm c.c.}$
\\
\hline
\end{tabular}
\end{center}
\end{table}
\begin{table}[h]
\setlength{\extrarowheight}{5pt}
\begin{center}
\caption[Table caption]{Event samples used in the analysis.  For each
sample, the corresponding reactions from Table \ref{table:decay_modes}
are indicated.  In the tagged samples, the D$^{0}$ is assumed to
originate from D$^{*}$ decay and the final state of the D$^{0}$ decay
is indicated by the subscript where `K$_{\rm sub}$' stands for a kaon
with momentum below the RICH threshold. Throughout this paper, each
sample will be referred to using the above notation.  \label{table:channels}}
\begin{tabular}{|c||c||c|c|c|c|}
\hline
\multirow{2}{*}{} & {\textbf{Untagged sample}} &\multicolumn{4}  {|c|}
  {\textbf{{Tagged samples} }}
\\ \hline
\textbf{Sample} & D$^{0}_{{\rm K}\pi}$ & D$^{*}_{{\rm K}\pi}$ & D$^{*}_{{{\rm K}_{\rm sub}}\pi}$ & D$^{*}_{{\rm K}\pi\pi^{0}}$ & D$^{*}_{{\rm K}\pi\pi\pi}$  \\ \hline
\textbf{Reaction number} & 1 & 1 & 1 & 2 & 3  \\ \hline
\end{tabular}
\end{center}
\end{table}
\begin{table}[h]
\setlength{\extrarowheight}{6pt}
\begin{center}
\caption{List of kinematic cuts used for each data sample. 
A D$^{0}$ candidate is accepted if it fullfils 
all conditions in a corresponding column. Here $\Delta M = 
M_{{\rm K}\pi\pi_{\rm s}}^{\rm rec} - M_{{\rm K}\pi}^{\rm rec} 
- M_{\pi}$, where the superscript `rec' denotes the reconstructed mass.
 \label{table:kin_cuts}}
\begin{tabular}{|c||c||c|c|c|c|}
\hline
   & \multicolumn{5}{c|}{\textbf{Kinematic cut intervals}} \\ \hline
 \multirow{2}{*}
{\textbf{Variables}} & {\textbf{Untagged sample}}  & \multicolumn{4}{|c|}{\textbf{Tagged samples}}  \\ 
\cline{2-2} \cline{3-3} \cline{4-4} \cline{5-5} \cline{6-6}
              &D$^{0}_{{\rm K}\pi}$& \textbf{D$^{*}_{{\rm K}\pi}$} & \textbf{D$^{*}_{{\rm K}\pi\pi^{0}}$} & 
\textbf{D$^{*}_{{\rm K}_{sub}\pi}$} & \textbf{D$^{*}_{{\rm K}\pi\pi\pi}$} \\ \hline 
    \textbf{$(M_{{\rm K}\pi}^{\rm rec} - M_{{\rm D}^{0}})\ [{\rm MeV}/c^{2}]$} & 
$ [-400,\ +400]$ & \multicolumn{2}{|c|}{$ [-600,\ +600]$}  &  \multicolumn{2}{|c|}{$
[-400,\ +400]$}   \\ \hline
    \textbf{$|\cos\theta^{*}|$} & $< 0.65$ & \multicolumn{2}{|c|}{$< 0.90$} & 
\multicolumn{2}{|c|}{$< 0.85$}   \\ \hline  
    \textbf{$z_{{\rm D}^{0}}$} &  $ [0.20,\ 0.85]$ & \multicolumn{2}{|c|}{$ [0.20,\ 0.85]$} & 
$ [0.25,\ 0.85]$  & $ [0.30,\ 0.85]$  \\ \hline
    \textbf{${ p}_{\rm K}\ [{\rm GeV}/c]$} & 
$\ [9.5,\ 50]$  &\multicolumn{2}{c|}{$\ [9.5,\ 50]$}  & $ [2.5,\ 9.5]$ & $\ [9.5,\ 
50]$  \\  \hline
         \textbf{${ p}_{\pi}\ [{\rm GeV}/c]$}  
&  $ [7,\ 50]$ & \multicolumn{4}{c|}{$ [2.5,\ 50]$} \\ \hline
    \textbf{{$\Delta M\ [{\rm MeV}/c^{2}]$}} 
& \textbf{-----} & \multicolumn{3}{|c|}{$ [3.2, 8.9]$}  & $ [4.0, 7.5]$   \\ \hline
    \textbf{{$p_{\pi_{s}} [{\rm GeV}/c]$}} 
& \textbf{-----} & \multicolumn{4}{|c|}{$ < 8$}   \\ \hline
 \end{tabular}
\end{center}
\end{table}

Particles are identified using the RICH detector.  Using the measured
momentum of a charged particle and the distributions of Cherenkov
photons, likelihood values for different mass hypotheses and for the
background hypothesis are computed.  A particle is identified as kaon
or pion if the likelihood value is larger than those for all remaining
hypotheses.  This procedure is very efficient in reducing the
combinatorial background of two particles other than $\pi$ and K.  A
detailed description of the identification procedure is given in
Ref. \cite{celso_phd}.

\begin{figure}[!h]
\begin{center}
\includegraphics[totalheight=6cm,width=0.6\textwidth]{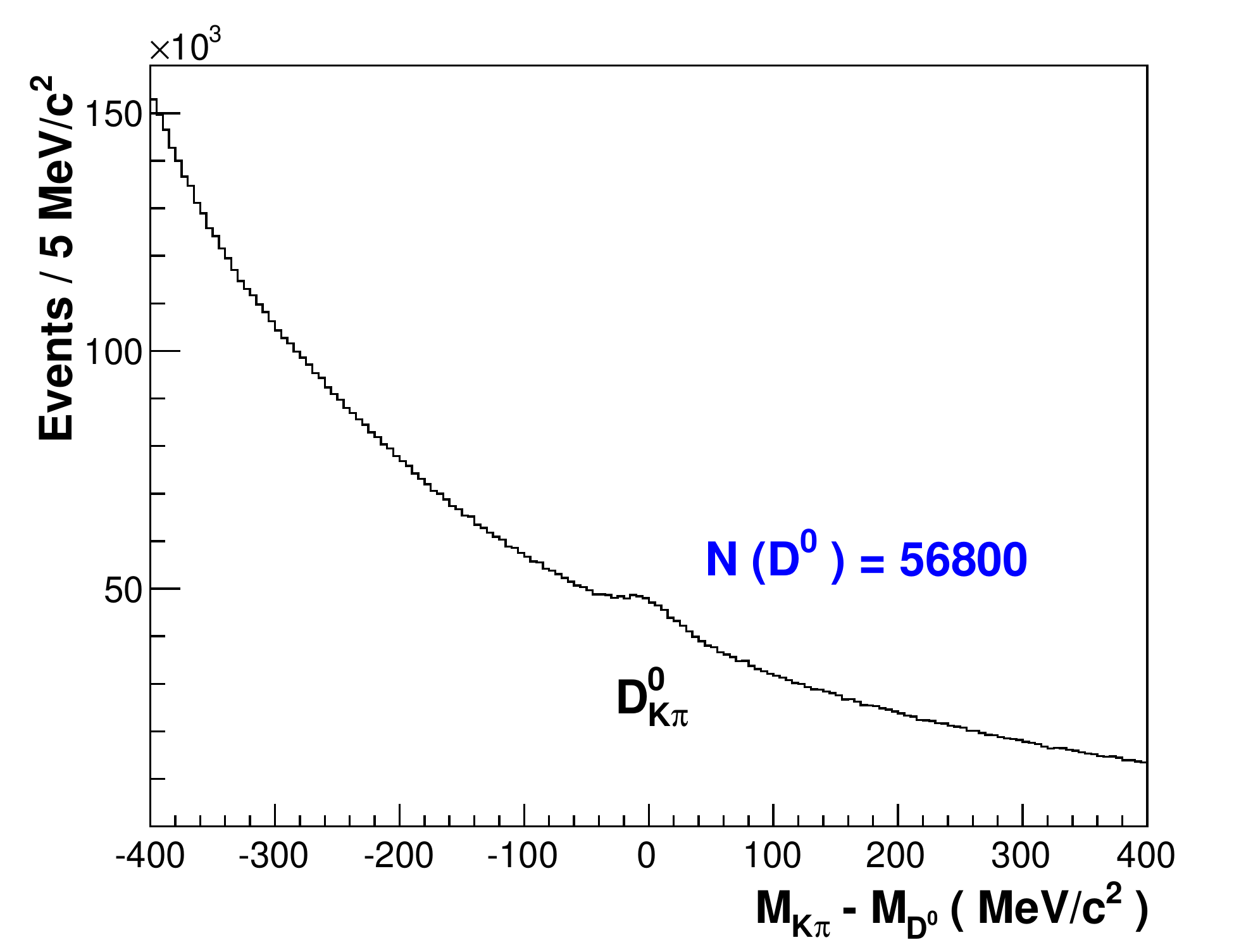}
\includegraphics[totalheight=6cm,width=0.6\textwidth]{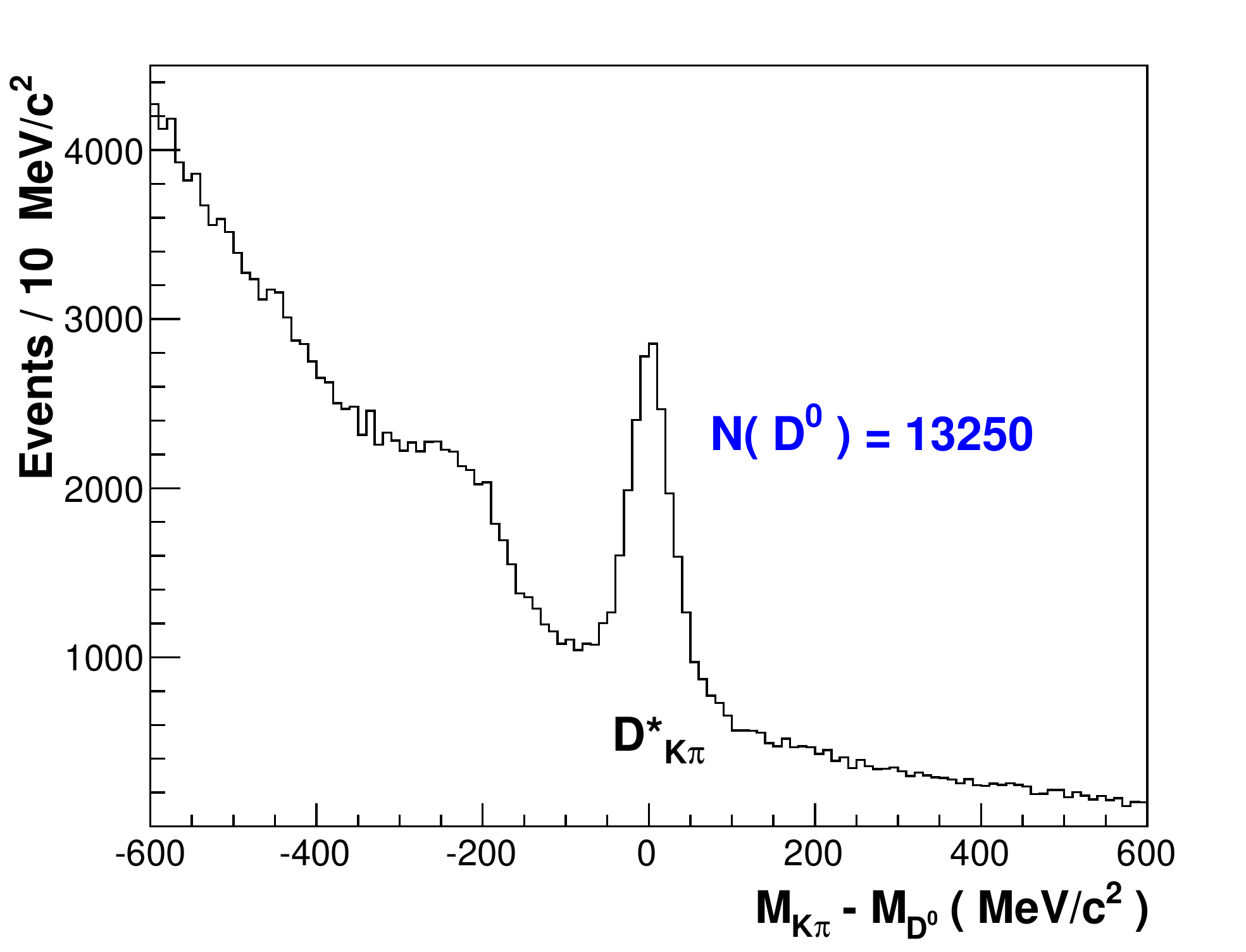}
\caption{Invariant mass spectra for the D$^0_{{\rm K}\pi}$ and 
D$^*_{{\rm K}\pi}$ samples with the  approximate number of D$^0$ mesons 
above background.
}
\label{fig:masses_for_all_samples_ab}
\end{center}
\end{figure}

The following selection criteria were applied to obtain the final
event samples.  The untagged sample D$^0_{{\rm K}\pi}$ contains events
with K$\pi$ pairs in the reconstructed mass range given in Table
\ref{table:kin_cuts}, which do not stem from decays of reconstructed
D$^*$ mesons, see Fig. \ref{fig:masses_for_all_samples_ab}.  Due to
large combinatorial background this sample requires more restrictive
cuts for the identification of pion and kaon: a pion momentum above 7\
GeV/$c$ is required to avoid contamination from electrons.  For the
four tagged samples, a D$^*$ meson is selected by requiring the
presence of a slow pion, $p_{\pi_{\rm s}} < $ 8 GeV/$c$, in addition
to a D$^0$ candidate.  The presence of the slow pion permits the
application of two additional cuts. The first one uses the RICH
detector to reject electrons that mimic slow pion candidates and
reduces the combinatorial background by a factor of two. The second
one is a cut on the mass difference, $\Delta M = M_{{\rm K}\pi\pi_{\rm
s}}^{\rm rec} - M_{{\rm K}\pi}^{\rm rec} - M_{\pi}$, where $M_{{\rm
K}\pi\pi_{\rm s}}^{\rm rec}$ and $M_{{\rm K}\pi}^{\rm rec}$ are the
reconstructed masses of the D$^{*}$ and the D$^{0}$ candidates,
respectively.  This mass difference can be measured with very good
precision and thus the cut on $\Delta M$ results in a significant
reduction of the combinatorial background in the tagged samples.

\begin{figure}[!h]
\begin{center}
\begin{minipage}[c]{0.46\linewidth}
\includegraphics[totalheight=7.5cm,width=1.0\textwidth]{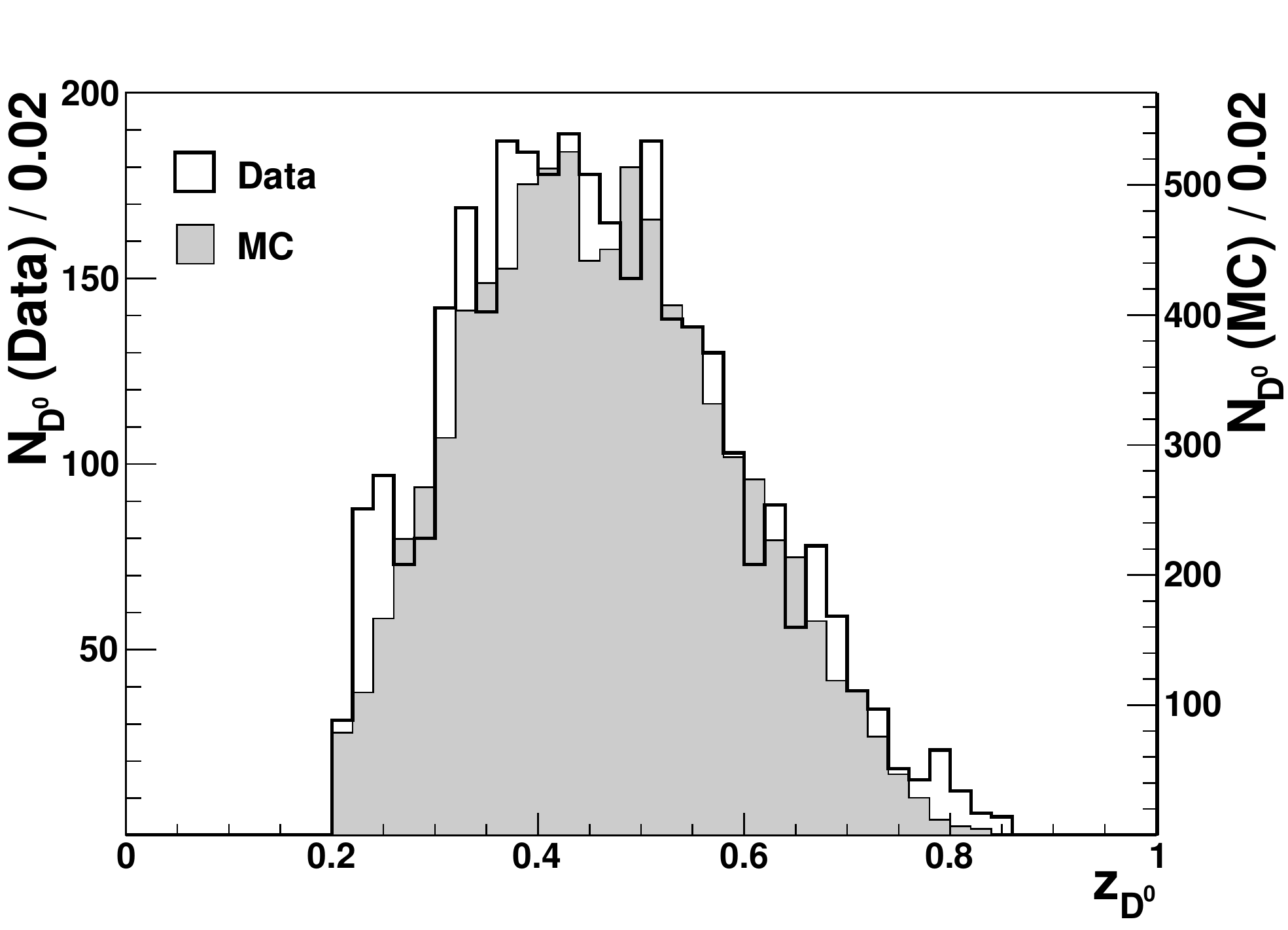}
\end{minipage}
\end{center}
\caption{Distribution of $z_{{\rm D}^{0}}$ for the 
 D$^*_{{\rm K}\pi}$ data sample (background is subtracted) and  
corresponding Monte Carlo events. D$^0$ mesons are selected in the
$\pm 80$ MeV/$c^2$ mass window around the D$^0$ mass.}
\label{fig:ZD_MC_vs_Data}
\end{figure}

In addition to the cuts described above, further kinematic cuts were
applied to all samples. It is demanded that
$|\mathrm{cos}\theta^*|<0.9$ for the tagged D$^*_{{\rm K}\pi}$ and
D$^*_{{\rm K}\pi\pi^0}$ samples, $|\mathrm{cos}\theta^*|<0.65$ for the
sample D$^0_{{\rm K}\pi}$ and $|\mathrm{cos}\theta^*|<0.85$ for the
remaining samples.  This cut suppresses mainly background events and
improves the significance of the signal.  Finally, all events have to
satisfy a cut on $z_{{\rm D}^{0}}$.  Since a pair of charmed quarks is
produced in the centre-of-mass of the $\gamma^{*}g$ system, each quark
receives on average half of the virtual photon energy. Indeed, the
measured $z_{{\rm D}^{0}}$ distribution and the one simulated assuming
a pure PGF process (with parton showers included) are very similar and
have a most probable value close to 0.5, see
Fig. \ref{fig:ZD_MC_vs_Data}.  This fact strongly supports the
assumption on PGF dominance in charm production.

\vspace{0.2 cm}
\begin{figure}[!h]
\begin{center}
\includegraphics[totalheight=6cm,width=0.6\textwidth]{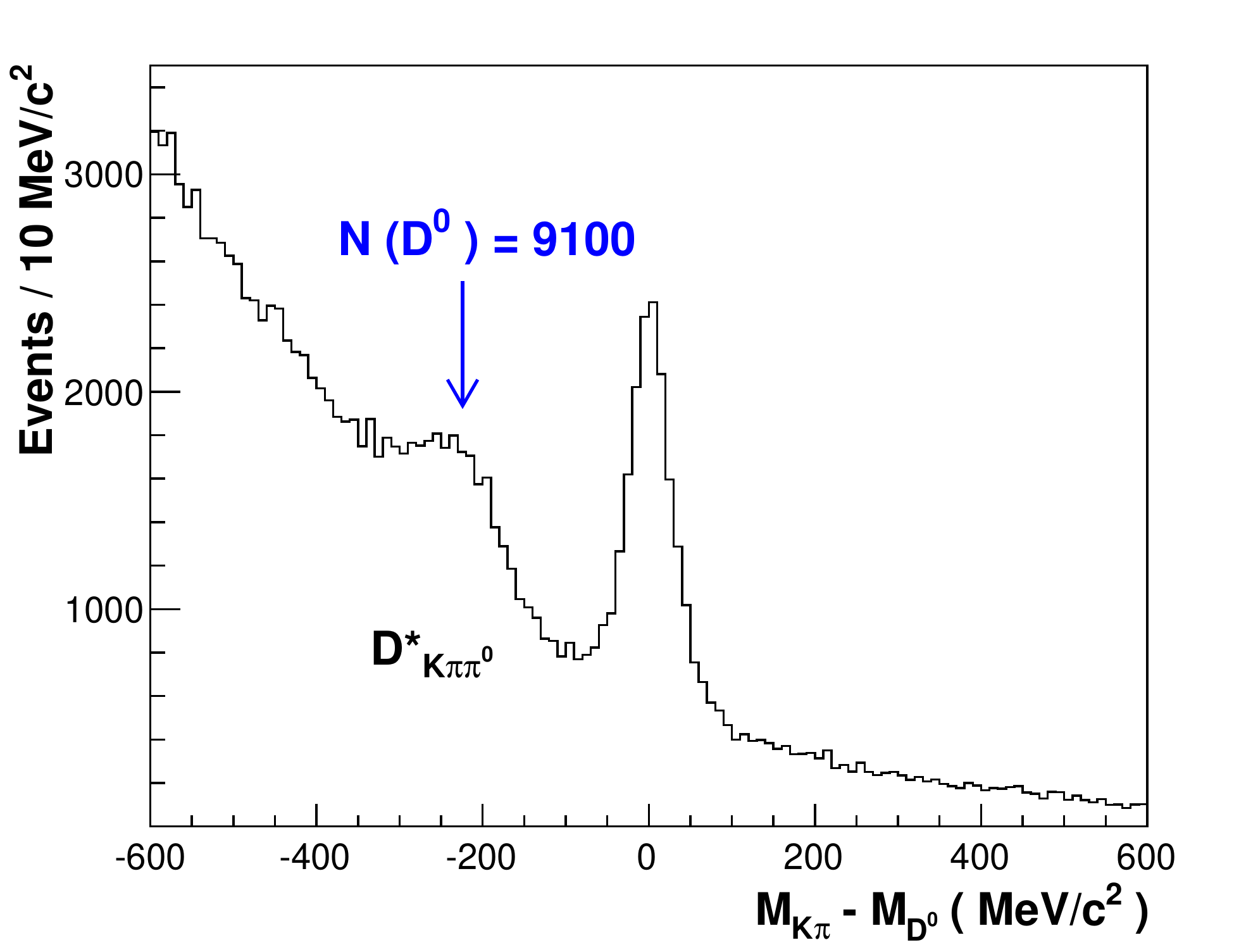}
\includegraphics[totalheight=6cm,width=0.6\textwidth]{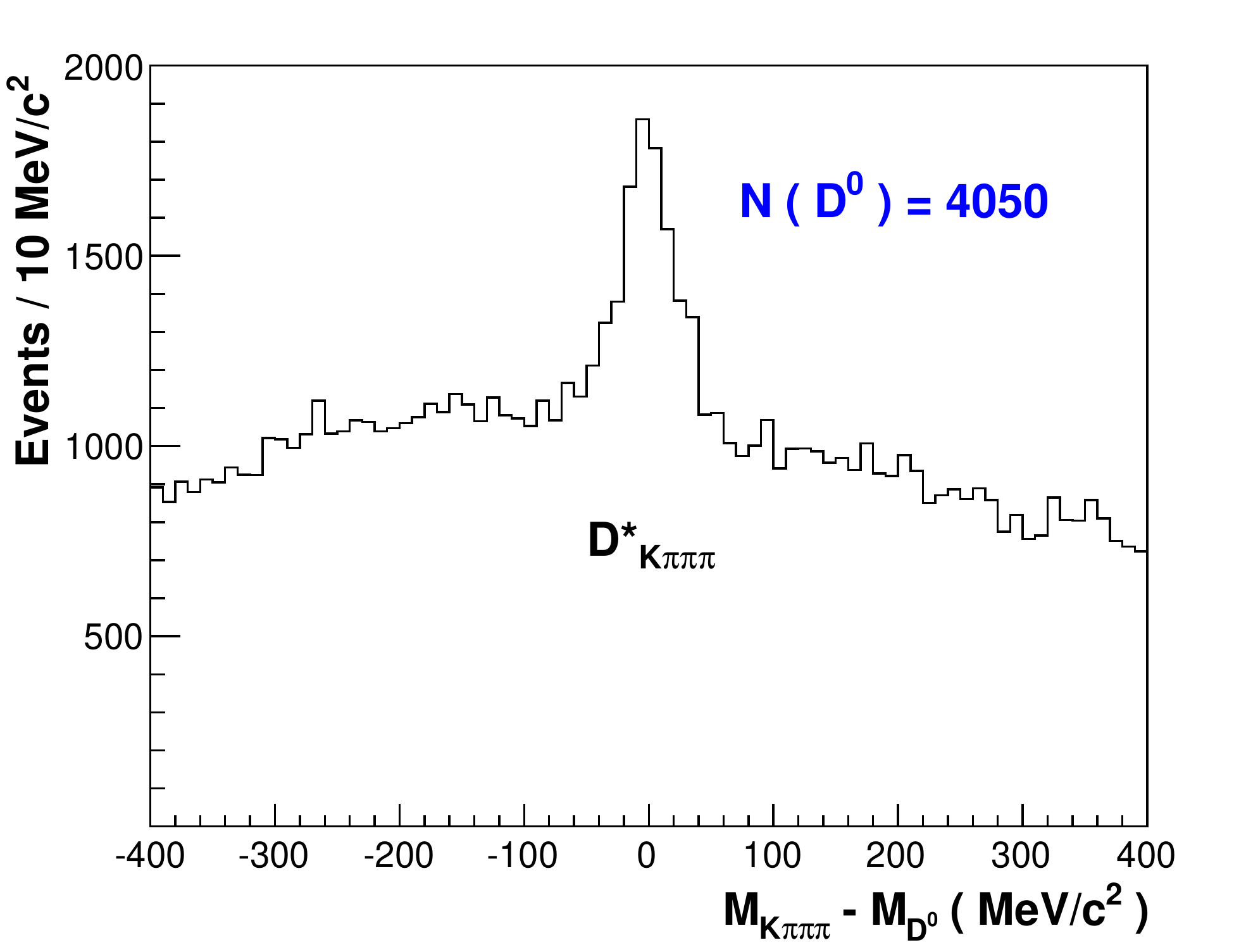}
\includegraphics[totalheight=6cm,width=0.6\textwidth]{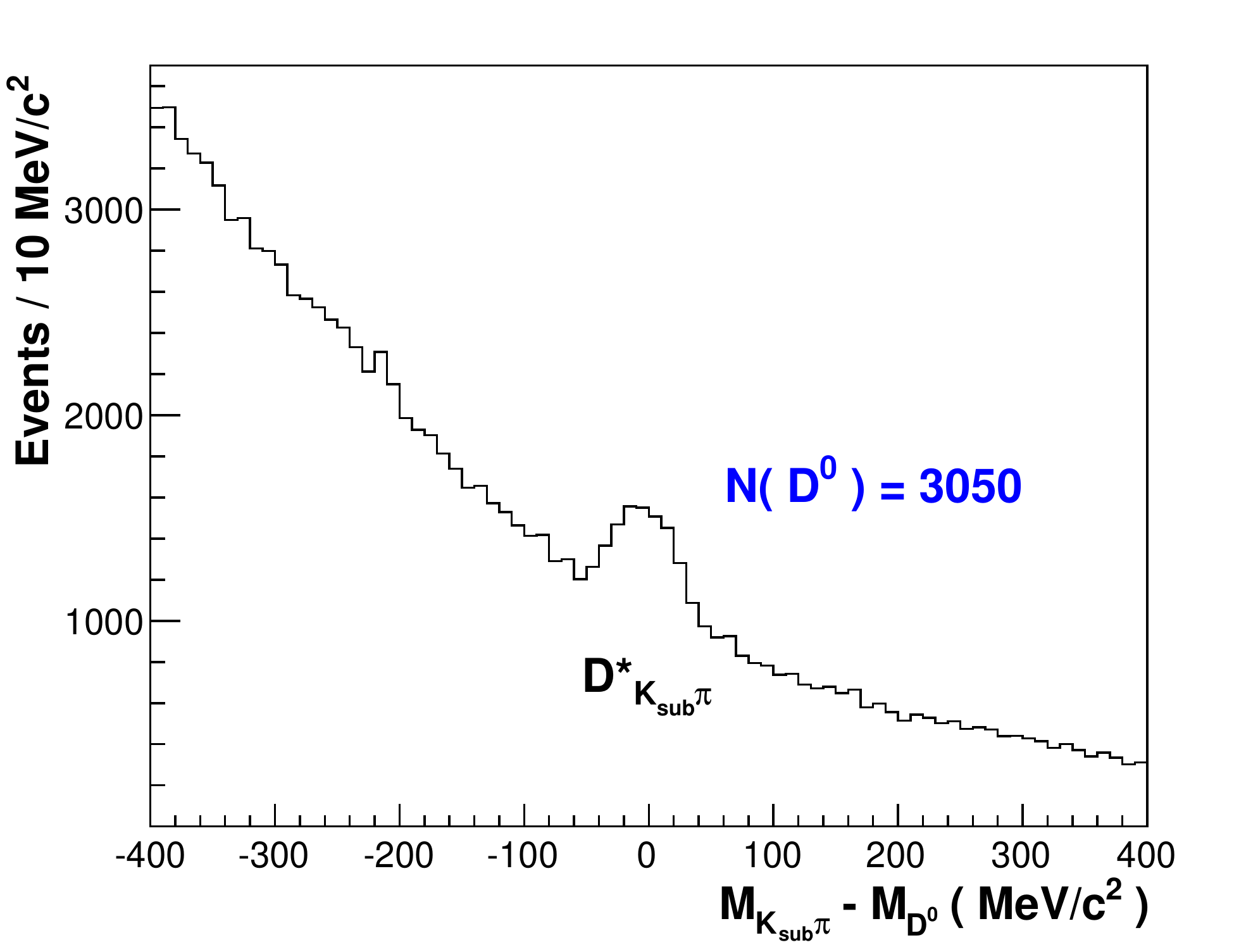}
\caption{Invariant mass spectra for the D$^*_{{\rm K}\pi\pi^0}$,
D$^*_{{\rm K}\pi\pi\pi}$ and D$^*_{{\rm K_{sub}}\pi}$ samples. 
The purity of the
samples was improved using the Neural Network.
The approximate number of D$^0$ mesons above background is given.
}
\label{fig:masses_for_all_samples_abc}
\end{center}
\end{figure}

Final mass spectra for the untagged and tagged samples, selected
according to kinematic cuts listed in Table \ref{table:kin_cuts}, are
shown in Fig.  \ref{fig:masses_for_all_samples_ab}.  For the latter, a
most pronounced signal is visible at the D$^0$ mass with a mass
resolution of about 27~MeV/$c^2$.  In this mass spectrum, also a
second structure is visible at about \mbox{--250 MeV/$c^2$} which is
due to events with ${\rm D}^{0} \ {\longrightarrow} \ {\rm
K}^{-}\pi^{+}\pi^{0} \ $ decays with neutral pion not reconstructed in
the analysis. Thus the K$\pi$ spectrum is shifted to lower mass as
compared to ${\rm D}^{0} \ {\longrightarrow} \ {\rm K}^{-}\pi^{+} \ $
decays.  The purity of this signal is much worse due to the
non-reconstructed neutral pion.

Further improvement of the significance of the signal is accomplished
by applying the Neural Network method described in Sec. \ref{sec:ssb}
which leads to a considerable reduction of the combinatorial
background in the tagged samples.  The resulting mass spectrum for
D$^*_{{\rm K}\pi\pi^0}$ is shown in Fig.
\ref{fig:masses_for_all_samples_abc}, with an improvement of the
signal strength by 15\% while for the ${\rm D}^{0} \ {\longrightarrow}
\ {\rm K}^{-}\pi^{+} \ $ the signal and the background are reduced in
a similar way so that the significance of the signal stays
unchanged. Therefore only the criteria from Table~\ref{table:kin_cuts}
are used to select the final D$^*_{{\rm K}\pi}$ sample, see
Fig.~\ref{fig:masses_for_all_samples_ab}.

Results on channels with a weaker D$^0$ signal like D$^*_{{\rm
K}\pi\pi\pi}$ and D$^*_{{\rm K_{sub}}\pi}$, are also shown in
Fig. \ref{fig:masses_for_all_samples_abc}.  The sample D$^*_{{\rm
K_{sub}}\pi}$ contains events where the momentum of the kaon
candidate is below the limit of 9 GeV/$c$ for kaon identification by
the RICH detector.  Simulations using a Monte Carlo generator for
heavy flavours, AROMA \cite{aroma}, and a full spectrometer
description based on GEANT \cite{geant} have shown that about 30\% of
the kaons coming from D$^{0}$ decays have their momenta below this
RICH threshold.  Therefore it is only required that those particles,
K$_{\rm sub}$, are not identified as pions or electrons.

In the case the two D$^{0}$ candidates are found in the same event,
only one of them, chosen randomly, is considered in the analysis.  If
two channels contribute with a D$^{0}$ candidate to the same event,
only one of them is accepted according to the following priority rule:
D$^*_{{\rm K}\pi\pi\pi}$, D$^*_{{\rm K}\pi}$ or D$^*_{{\rm
K}\pi\pi^0}$, D$^0_{{\rm K}\pi}$, D$^*_{{\rm K_{sub}}\pi}$ (see
Ref. \cite{celso_phd}).

\begin{figure}[!h]
\begin{center}
\includegraphics[totalheight=6.5cm,width=0.9\textwidth]{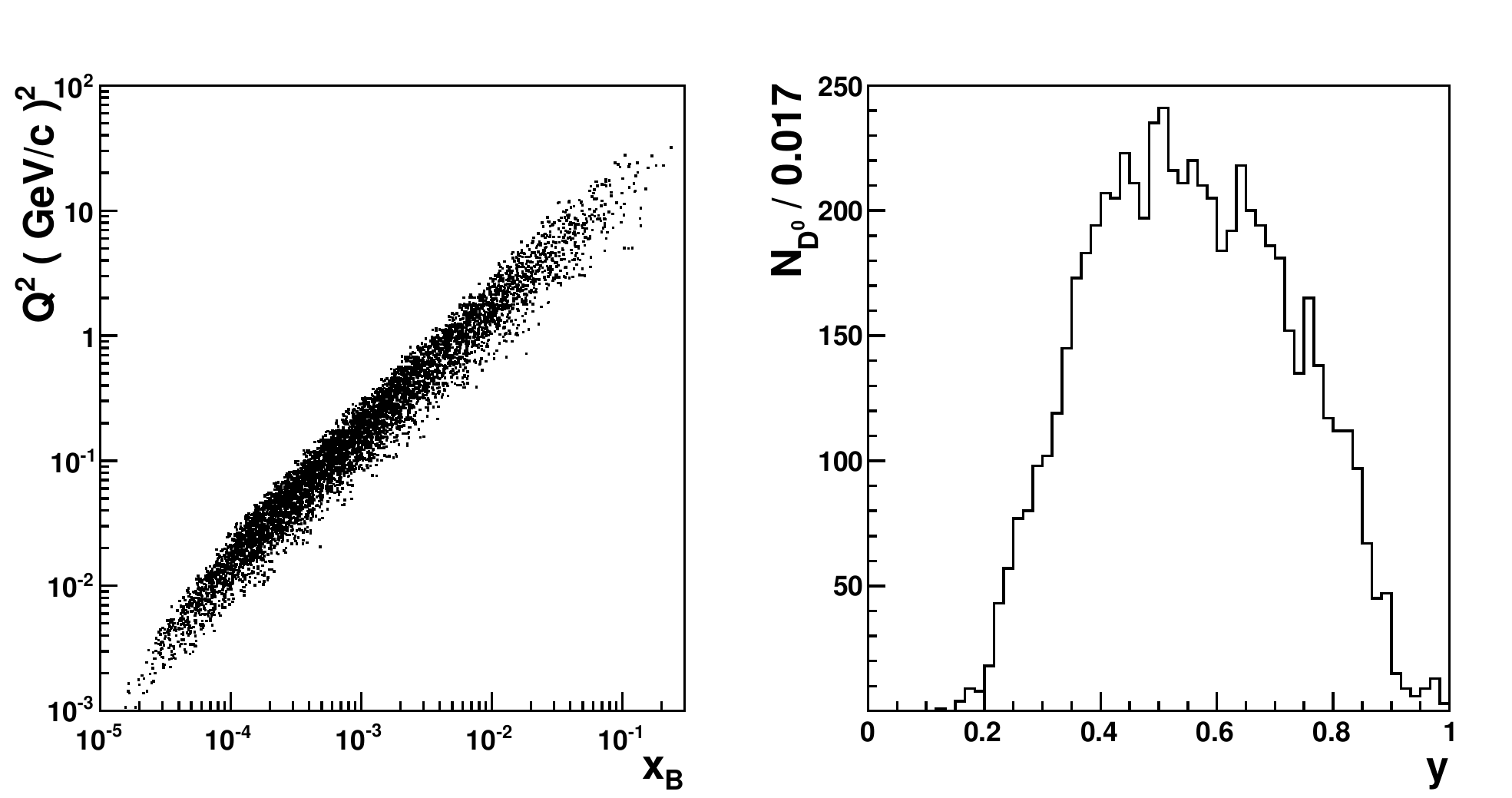}
\caption{A scatter plot of $Q^2$ {\it vs} $x_{\rm B}$ and a distribution 
of $y$ for the 2006 D$^*_{{\rm K}\pi}$
sample selected as in Fig. \ref{fig:ZD_MC_vs_Data} except that the
background is not subtracted.}
\label{fig:QXY_distribution}
\end{center}
\end{figure}

Distributions of $x_B$, $Q^2$ and $y$ variables for the D$^*_{{\rm
K}\pi}$ candidates from 2006, and from the D$^0$ signal region, are
presented in Fig. ~\ref{fig:QXY_distribution}. For those events
$x_{\rm B}$ values range from about $10^{-5}$ to 0.1 with $\langle
x_{\rm B}\rangle$ = 0.004, $Q^{2}$ values from 10$^{-3}$ to 30
(GeV/$c$)$^{2}$ with $\langle Q^{2}\rangle$ = 0.6 (GeV/$c$)$^{2}$, and
$y$ values from 0.1 to 1 with $\langle y \rangle$ = 0.63.

\section{Asymmetry evaluation}
\label{sec:Method}
In this section, we describe the determination of the virtual photon
asymmetry for D$^0$ production, $A^{\gamma {\rm N}} \equiv A^{\gamma
{\rm N} \rightarrow {\rm D}^0 {\rm X}}$, from the event samples
defined in Table \ref{table:channels}.  The method is similar to the
one used in our previous publication \cite{compass_pgf}. The asymmetry
$A^{\gamma {\rm N}}$ can be used in various ways to evaluate the gluon
polarisation $\left< \Delta g/g \right>$ at LO or NLO QCD accuracy.

\subsection{Analysis method}
\subsubsection{Asymmetries}
\label{Asymmetries}
The number of events collected  in a given target cell and
time interval is given by:
\begin{equation}
\label{yield}
 \frac{{\rm d}^k N}{{\rm d}m\, {\rm d}X}= a \phi n (s+b)  
  \left [1 + P_{\rm t} P_{\mu} f \left(
          \frac{s}{s+b}  A^{\mu {\rm N}\rightarrow \mu' {\rm D}^0 \rm{X}}+ 
         \frac{b}{s+b}
  A_{\rm B} \right)\right ]\,\,.
\end{equation}
Here, $A^{\mu {\rm N}\rightarrow \mu' {\rm D}^0 {\rm X}}$ is the
longitudinal double spin asymmetry of the differential cross section
for events with a D$^0$ or $\overline{{\rm D}^0}$ in the final state,
and $A_{\rm B}$ is the corresponding asymmetry originating from
background events.  Furthermore, $m\equiv M_{{\rm K}\pi}$ (or $m\equiv
M_{{\rm K}\pi\pi\pi}$) and the symbol $X$ denotes a set of $k-1$
kinematic variables describing an event ($p_{\rm T}^{\rm D^0}$,
$E_{\rm D^0}$, $Q^2$, $y$, \ldots), while $a$, $\phi$ and $n$ are the
spectrometer acceptance, the incident muon flux integrated over the
time interval, and the number of target nucleons respectively.  The
differential unpolarised cross sections for signal and background,
folded with the experimental resolution as a function of $m$ and $X$,
are represented by $s=s(m,X)$ and $b=b(m,X)$ respectively.  The ratio
${s}/(s+b)$ will be called `signal purity' and the ratio ${b}/(s+b)$
`background purity'.  The information on the gluon polarisation is
contained in the virtual photon asymmetry $A^{\gamma {\rm N}} = A^{\mu
{\rm N}\rightarrow \mu' {\rm D}^0{\rm X}}/D$.  Similarly, the
background asymmetry can be written as $A^{\gamma {\rm N}}_{\rm B} =
A_{\rm B}/D$.  Here, $D$ is the so-called depolarisation factor
accounting for the polarisation transfer from lepton to virtual
photon:
\begin{equation}
\displaystyle
D = \frac{y  \left [ 2-y - 
\frac{2y^2m_\mu^2}{Q^2}\right ]} 
    { y^2 \left (1-\frac{2m_\mu^2}{Q^2}\right )  + 2(1-y)}.
\label{depolarisation_fac}
\end{equation}

A straightforward way to extract $A^{\gamma {\rm N}}$ would be the
following: Eq.~(\ref{yield}) is integrated over the variables $X$ to
obtain the number of events in both spin configurations as a function
of the invariant mass $m$. Next, the event number asymmetry in the
${\rm D}^0$ signal region is extracted, and a possible background
asymmetry, determined from the asymmetries in sidebands to the left
and right from the signal region, is subtracted.

For this analysis, however, we choose the method of event weighting,
which is advantageous in terms of statistical precision.  Compared to
previous COMPASS analyses where weighting procedures were
applied~\cite{compass}, here the weighting procedure is extended to
determine the background asymmetry $A_{\rm B}^{\gamma{\rm N}}$
simultaneously with $A^{\gamma{\rm N}}$ \cite{pretz}. In order to
achieve this, every event is weighted once with a signal weight,
$w_{\rm S}$, and once with a background weight, $w_{\rm B}$:
\begin{eqnarray}
w_{\rm S} &=& P_{\mu} f D \frac{s}{s+b} \, ,\label{weight_ws}\\ w_{\rm
B} &=& P_{\mu} f D \frac{b}{s+b} \, . \label{weight_wb}
\end{eqnarray}
Except for the target polarisation $P_{\rm t}$, these weights are the
prefactors of the asymmetries $A^{\gamma{\rm N}}$ and $A_{\rm
B}^{\gamma{\rm N}}$, see Eq. (\ref{yield}). The target polarisation is
not included in the weights because its time dependence would lead to
an increase in false asymmetries.  The signal and background purities
are included in the respective weights.  This procedure leads to the
highest possible statistical precision which would also be obtained in
the unbinned maximum likelihood method~\cite{pretz_hab,pretz}.  Note
that the unbinned maximum likelihood method, cannot be applied here
because the acceptance and flux factors in Eq. (\ref{yield}) are not
known with sufficient precision, only their ratios for different spin
states and target cells are known. These ratios will be used for the
extraction of $A^{\gamma N}$ and $A_{\rm B}^{\gamma N}$.

The asymmetry $A^{\gamma {\rm N}}$ is extracted in bins of D$^0$
transverse momentum with respect to the virtual photon, $p_{\rm
T}^{{\rm D}^0}$, and D$^0$ energy in the laboratory system, $E_{{\rm
D}^0}$.  These variables and their binning were chosen to minimise the
influence of the experimental acceptance on the asymmetry.  As
$A^{\gamma {\rm N} }$ does not contain the depolarisation factor $D$,
its remaining dependence on the inclusive variables $y$ and $Q^2$ is
very weak.  The expectation value of the sum of signal weights is
obtained as:

\begin{eqnarray}
  \left< \sum_{i=1}^{N_t} w_{{\rm S},i} \right> &=& \int w_{\rm S}(X,m) \,\frac{{\rm
      d}^k N_t}{{\rm d}m\, {\rm d}X} {\rm d}m {\rm d}X  \nonumber\\
&=&  \alpha_{{\rm S},t}
 \left[1 +  \left<  \beta_{\rm S}\right>_{w_{\rm S}} \left\langle A^{\gamma{\rm N}} \right\rangle_{w_{\rm S}\beta_{\rm S}} + 
         \left<  \beta_{\rm B}\right>_{w_{\rm S}} \left\langle A_{\rm B}^{\gamma N}\right\rangle_{w_{\rm S}\beta_{\rm B}} \right]  \label{sum_w}.
\end{eqnarray}
The symbols used are defined as:
\begin{eqnarray}
 \alpha_{{\rm S},t} &=& \int w_{\rm S}~a_t~\phi_t~n_t~(s+b)~{\rm d}m~{\rm d}X\, , \\ 
\left< \eta \right>_w &=& \frac{\int \eta w a_t \phi_t n_t (s+b) {\rm d}m  
{\rm d}X}{\int w a_t \phi_t n_t (s+b) {\rm d}m {\rm d}X}\, , 
\label{avg_beta_s}
 \end{eqnarray}
with $\beta_{\rm S} = w_{\rm S} P_{\rm t}$, $ \beta_{\rm B} = w_{\rm
B} P_{\rm t}$, $\eta \in \left [\beta_{\rm S}, \beta_{\rm B},
A^{\gamma{\rm N}}, A^{\gamma{\rm N}}_{\rm B}\right]$ and $w \in \left
[w_{\rm S}, w_{\rm B}, w_{\rm S}\beta_{\rm S}, w_{\rm S}\beta_{\rm
B}\right]$.  The index $t$ denotes the target cell before ($t=u,d$) or
after ($t=u',d'$) spin rotation and $N_t$ is the number of events
observed in a given target cell\footnote{ In 2006--2007 $d$ and $u$
stand for the central target cell and the sum of the outer ones,
respectively.  }.  An equation analogous to Eq. (\ref{sum_w}) holds
for the sum of background weights $\langle \sum_{i=1}^{N_t} w_{{\rm
B},i} \rangle$, with analogous definition of symbols.  In total, eight
equations similar to Eq. (\ref{sum_w}) are obtained for every ($p_{\rm
T}^{\rm D^0}, E_{\rm D^0}$) bin: for the signal and background weights
in two target cells and for two spin configurations. These eight
equations contain 12 unknowns: $\langle A^{\gamma{\rm
N}}\rangle_{w_{\rm S}\beta_{\rm S}}, \langle A^{\gamma{\rm N}}_{\rm
B}\rangle_{w_{\rm S}\beta_{\rm B}}, \langle A^{\gamma{\rm
N}}\rangle_{w_{\rm B}\beta_{\rm S}}, \langle A^{\gamma{\rm N}}_{\rm
B}\rangle_{w_{\rm B}\beta_{\rm B}}$, four acceptance factors
$\alpha_{\rm S,t}$ and four acceptance factors $\alpha_{\rm B,t}$.

The expectation values of the sum of weights on the left hand side of
Eq.~(\ref{sum_w}) are identified with the measured sums of weights.
In order to extract $A^{\gamma{\rm N}}$ and $A_{\rm B}^{\gamma {\rm
N}}$ from the measured sums of weights one proceeds as follows.  The
factors $\langle\beta_{\rm S,B}\rangle_{w_{\rm S},w_{\rm B}}$ are
evaluated from data, {\it e.g.}
\begin{equation}\label{sum_beta}
 \left< \beta_{\rm S}\right>_{w_{\rm S}} \approx
   \frac{\sum_{i=1}^{N_t} \beta_{\rm S} w_{\rm S}}{\sum_{i=1}^{N_t}
   w_{\rm S}} \, .
\end{equation}
The expectation values appearing in Eq.~(\ref{avg_beta_s}) contain the
spin-averaged cross section while the sum over events in
Eq.~(\ref{sum_beta}) used to evaluate these expectation values runs
over the spin-dependent events.  This has a negligible effect on the
result because the raw asymmetry $P_{\rm t} P_{\mu} f D \langle
A^{\gamma{\rm N}}\rangle$ is very small.  This smallness makes sure
that neither the result nor its statistical error are sensitive to the
fact that the same data which are used to determine the asymmetries
are also used to evaluate the expectation values above.

The acceptance factors $\alpha_{\rm{S},t}$ and $\alpha_{\rm{B},t}$
cannot be determined with sufficient precision to extract
$A^{\gamma{\rm N}}$ ~and $A_{\rm B}^{\gamma N}$ directly from the set
of eight equations. By assuming that for both signal and background
possible acceptance variations affect the upstream and downstream
cells in the same way, {\it i.e.}  ${\alpha_{\rm }^{u}/ \alpha_{\rm
}^{d}} = {\alpha_{\rm }^{u'}/ \alpha_{\rm }^{d'}}$, the number of
unknowns is reduced to ten.  With an extra, much weaker assumption
that signal and background events from the same target cell are
affected in the same way by the acceptance variations, one arrives at
a system of eight equations with nine unknowns.  Possible deviations
from the above assumptions may generate false asymmetries which are
included in the systematic uncertainty, see Sec. \ref{sec:results}.

The number of unknowns is reduced to seven with two additional
assumptions:
\begin{equation}
\left< A^{\gamma{\rm N}}\right>_{w_{\rm S}\beta_{\rm S}} =
\left< A^{\gamma{\rm N}}\right>_{w_{\rm B}\beta_{\rm S}} = A^{\gamma{\rm N}}
{\rm ~~~~and~~~~} \left< A^{\gamma{\rm N}}_{\rm B}\right>_{w_{\rm S}\beta_{\rm B}} =
\left< A^{\gamma{\rm N}}_{\rm B}\right>_{w_{\rm B}\beta_{\rm B}} = 
A^{\gamma{\rm N}}_{\rm B}, 
\label{reduction}
\end{equation}
which are satisfied for constant values of $A^{\gamma{\rm N}}$ and
$A^{\gamma{\rm N}}_{\rm B}$ in a given bin.  The uncertainty on the
gluon polarisation introduced by this assumption will be discussed in
Sec. \ref{sec:results_new}.  Using the set of eight equations, the
asymmetry $A^{\gamma{\rm N}}$ and the background asymmetry $A_{\rm
B}^{\gamma{\rm N}}$ are determined simultaneously by a standard least
square minimisation procedure, which takes into account the
statistical correlation between $\sum w_{\rm S}$ and $\sum w_{\rm B}$
in the same target cell.  The correlation factor $\mbox{cov}\left(\sum
w_{\rm S},\sum w_{\rm B}\right)$ is given in Ref. \cite{pretz_hab}.
The analysis is performed independently for each ($p_{\rm T}^{\rm
D^0}, E_{\rm D^0}$) bin and D meson decay channel.

For determinations of average values of kinematic variables in each
($p_{\rm T}^{\rm D^0}, E_{\rm D^0}$) bin, a weight equal to $w_{\rm
S}^2$ is used, in accordance with Eq. (\ref{sum_w}).

\begin{figure}[h]
\begin{center}
\includegraphics[width=0.7\hsize,clip]{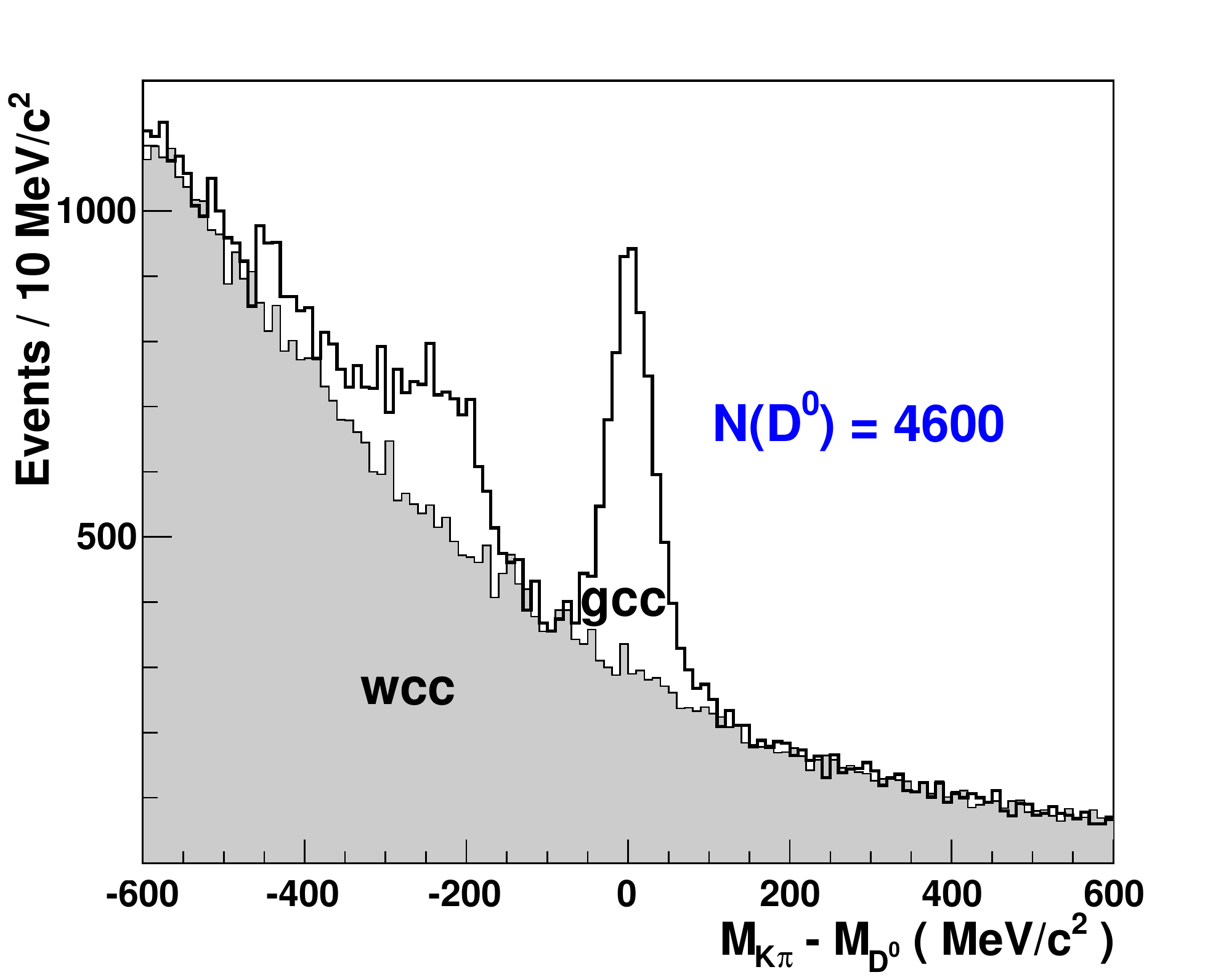}
\end{center}
\caption{
Example of the K$\pi$ invariant mass spectrum for the D$^*_{\rm K\pi}$
sample with good ({\it gcc}) and wrong ({\it wcc})
combination of pion and kaon charge signs. See text for details. 
Data were collected in 2007 with the \mbox{proton (NH$_3$)} target. 
}
\label{fig:celso3}
\end{figure}

\subsubsection{The signal purity}
\label{sec:ssb}
The signal purity $s/(s+b)$ can be extracted from a fit to the
invariant mass distribution of D$^0$ candidates.  It depends on
kinematic variables, for instance it is large at high transverse
momenta $p_{\rm T}^{\rm D^0}$ of the K$\pi$ system and small at low
$p_{\rm T}^{\rm D^0}$.  In order to implement the kinematic dependence
of the signal purity in the weights given by Eqs~(\ref{weight_ws},
\ref{weight_wb}), one would naively proceed by performing fits to the
corresponding invariant mass distributions in bins of kinematic
variables. This procedure is not feasible in our case because of
limited statistics.  Instead, in this analysis a classification based
on a Neural Network is employed \cite{celso_phd}.

Here, the aim of the Neural Network is to distinguish signal from
background events using only data. The network consists of information
processors (neurons), which are interconnected and organized into
layers.  The external information fed into the input layer is
processed in the hidden layers and the result produced by the output
layer is a classification of the event by the network.  In the present
case, the input layer contains a set of kinematic observables: ratios
of RICH likelihoods, $\cos\theta^*$, $z_D$ and kaon momentum.  There
are two hidden layers and the number of neurons in them varies during
the training process (dynamic network).  For each event, the network
tunes the strength of each variable-neuron and neuron-neuron
connection.  The strengths are obtained by minimising the squared
deviation between the expected output and the actual Neural Network
prediction.  This training process stops when the deviation reaches a
stable minimum \cite{neuralnet}.

\begin{figure}[htpb]
\begin{center}
\includegraphics[width=0.73\hsize,clip]{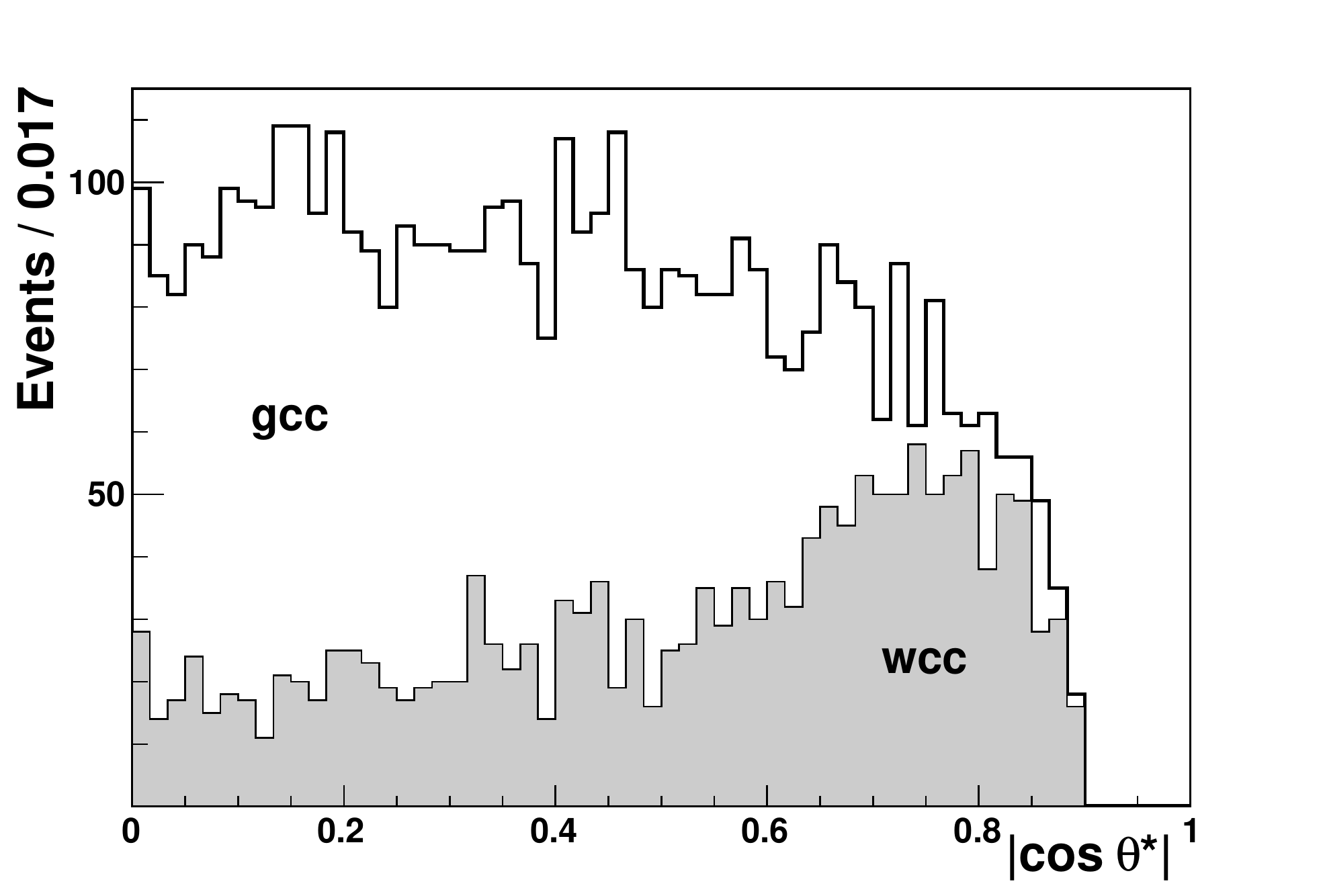}
\includegraphics[width=0.73\hsize,clip]{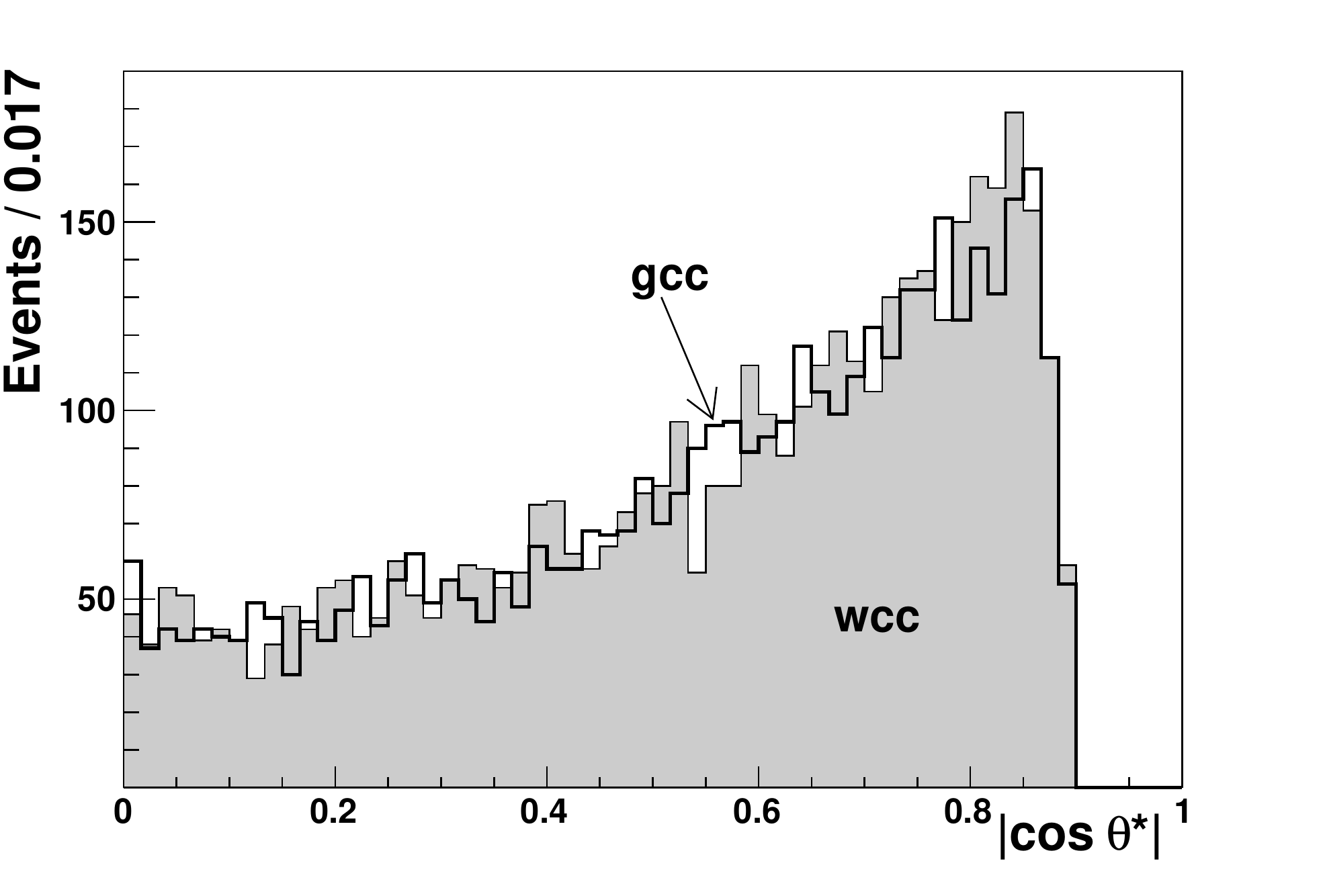}
\end{center}
\caption{
Example of the distribution of $\vert \cos\theta^*\vert$ (polar angle
of kaon in the D$^0$ centre-of-mass relative to the D$^0$ momentum) in
the D$^0$ meson rest frame for the {\it gcc} and {\it wcc} events
(D$^*_{{\rm K}\pi}$ sample, 2006 data). Top: region of the D$^0$
signal, bottom: outside the D$^0$ signal.
}
\label{fig:celso4}
\end{figure}

For each event sample (see Table \ref{table:channels}), two data sets
are used as inputs to the Neural Network. The first one contains the
D$^0$ signal and the combinatorial background events. These events are
called `good charge combination' ones ({\it gcc}) referring to the
charges of particles from D$^0$ decays, and they are selected as
described in Sec. \ref{sec:data_selection}.  The second set, the
`wrong charge combination' events ({\it wcc}), is selected in a
similar way except that the sum of charges of corresponding particles
should not be zero.  It contains only background events and is used as
a background model (see Fig. \ref{fig:celso3}). The Neural Network
performs a multi-dimensional comparison of {\it gcc} and {\it wcc}
events in a $\pm$40 MeV/$c^2$ mass window around the D$^0$
mass\footnote{A mass window of $\pm$30 MeV/$c^2$ is used for the
sample D$^0_{{\rm K}\pi}$ and $\pm$40 MeV/$c^2$ around --250 MeV/$c^2$
for the sample D$^*_{{\rm K}\pi\pi^0}$.}.  Within the {\it gcc} set,
signal events are distinguished from combinatorial background by
exploiting differences between the {\it gcc} and {\it wcc} sets in the
shapes of distributions of kinematic variables as well as
multi-dimensional correlations between them.  An example of a properly
chosen variable for the network is $\cos\theta^*$, as shown in
Fig. \ref{fig:celso4}.  The reconstructed mass cannot be used because
it would enhance the probability of a background event in the signal
region to be a true D meson.

The network classifies all the {\it gcc} events according to their
similarity in kinematics with respect to the {\it wcc} ones, and to
each event it assigns a probability of being a signal.  A probability
of 0.5 is assigned to indistinguishable events.  If the network is
trained with proper input samples, {\it i.e.} a correct background
model and a sufficiently strong signal, the network output,
${[s/(s+b)]}_{\rm NN}$, can be directly interpreted as an estimate of
the signal purity in the corresponding mass window.  This is the so
called `pure' Neural Network method, applicable to the D$^*_{{\rm
K}\pi\pi^{0}}$, D$^*_{{\rm K}\pi}$ and D$^*_{{\rm K}\pi\pi\pi}$
samples collected in 2004--2007, where event statistics and signal
purities are large.

The mass dependence of signal and background strengths, $s(m)$ and
$b(m)$, which cannot be obtained from the Neural Network in an
unbiased way, is determined from a fit to the mass spectra in bins of
${[s/(s+b)]}_{\rm NN}$. In order to describe the signal a Gaussian
distribution is used for all samples, while for the background the
following fitting functions are employed: two exponential
distributions for the D$^{0}_{{\rm K}\pi}$ channel and one exponential
for the D$^{*}$ tagged channels. An exception is the D$^*_{{\rm
K}\pi\pi\pi}$ sample, for which a second degree polynomial is used.
From those fits, corrections $\lambda$ to the signal purity are
obtained in the mass windows defined above:

\begin{equation}
\lambda = {\left\langle\left (\frac{s}{b}\right )_{\rm NN}
\right\rangle}~\frac{\displaystyle
\int_{}^{} b(m)
{\rm d}m} {\displaystyle\int_{}^{} s(m){\rm d}m}~,
\label{eq:lambda}
\end{equation}
\noindent
so that
\begin{equation}
\frac{s}{s+b} = \frac{\lambda s(m)}{\lambda s(m) +  b(m)}.
\end{equation}

\begin{figure}[htpb]
\begin{center}
\includegraphics[width=1.06\textwidth]{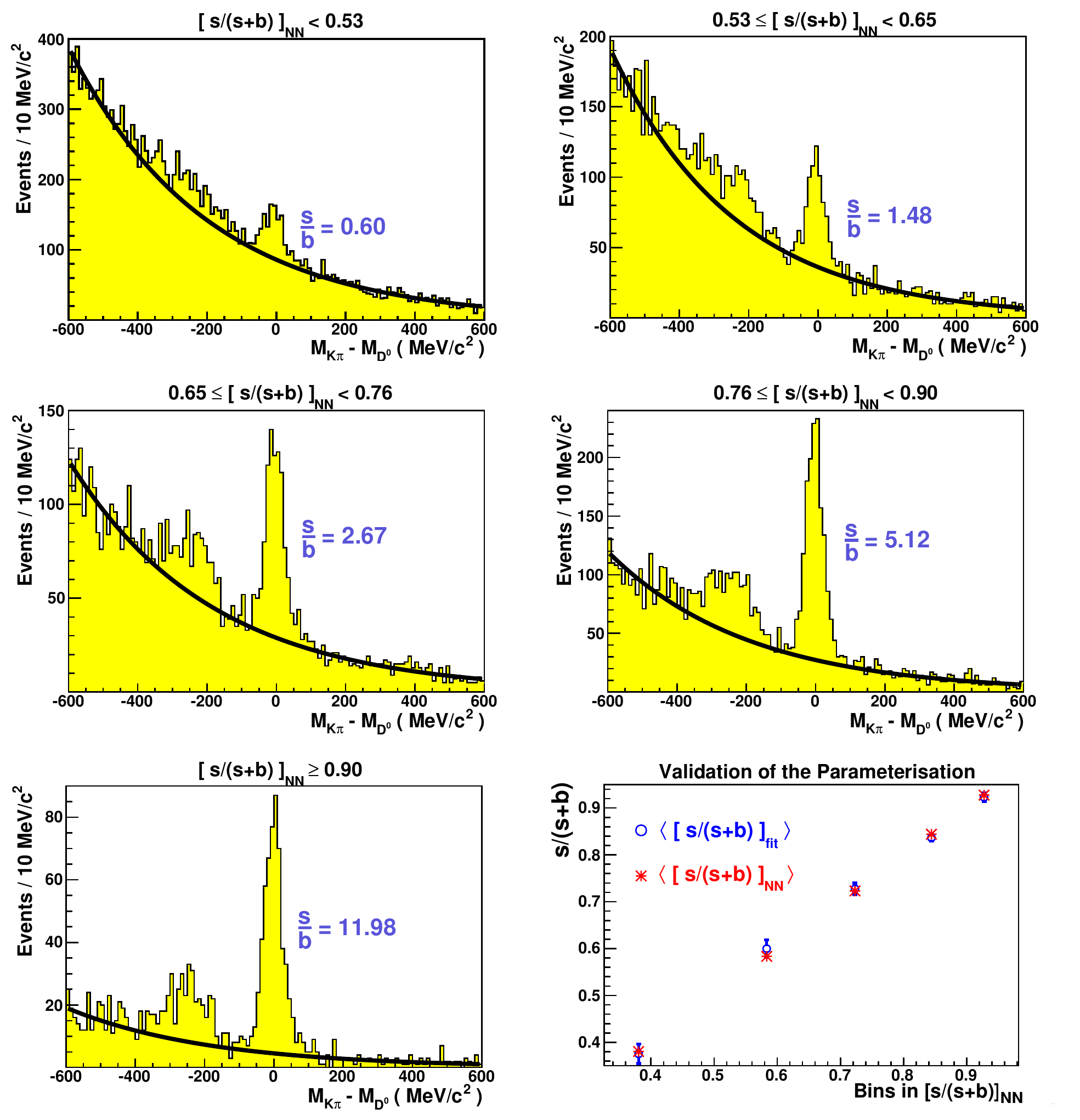}
\caption{The K$\pi$ invariant mass spectra in bins of the NN signal 
purity $[s/(s+b)]_{\rm NN}$ for the D$^*_{{\rm K}\pi}$ sample. 
The last panel shows a comparison of the two purities,
$[s/(s+b)]_{\rm NN}$ and $[s/(s+b)]_{\rm fit}$ (see text for details).
Curves show the background component of the invariant mass fits
described in the text. The significance of the D$^*_{{\rm K}\pi}$ signal 
is shown as ratio $s/b$.
}
\label{fig:celso2} 
\end{center}
\end{figure}

\noindent
The fit of the invariant mass spectra in bins of the NN signal purity
can also be used to validate the classification obtained by the Neural
Network.  For each bin, the signal purity is determined from an
integration of the signal and background fits over the mass windows
used.  Good agreement between signal purities from the NN and the fit
is found for all samples, which confirms that the Neural Network does
not introduce any bias in the analysis.  As an illustration, the mass
spectra in bins of the NN signal purity together with a comparison of
the two signal purities are shown for the D$^*_{{\rm K}\pi}$ sample in
Fig.~\ref{fig:celso2}. The signal purity clearly increases with
increasing $[s/(s+b)]_{\rm NN}$. Equally good agreement is found when
comparing $s/(s+b)$ between NN and data in bins of ($p_{\rm T}^{\rm
D^0}$, $E_{\rm D^0}$).

The signal purity can be parameterised in various ways provided it
correctly reproduces the data. Several parameterisations were found to
indeed yield asymmetries compatible within statistical
uncertainties. In order to achieve the statistically most precise
result on the gluon polarisation, we chose the method described in
Sec. \ref{sec:results_new}.  As a special feature in this method the
signal weight contains the product of the signal purity and $a_{\rm
LL}$, which is the partonic asymmetry for the PGF process.
Observation shows that $a_{\rm LL}$ is strongly anticorrelated with
the signal purity. Therefore a parameterisation of $a_{\rm LL}$,
validated as described in Ref. \cite{celso_phd}, was additionally
included in the training of the NN.

For the low purity sample D$^0_{{\rm K}\pi}$ collected in 2002--2007,
and for all samples collected in 2002 and 2003, the extraction of the
signal purities from the network is more complicated since the
anticorrelation mentioned above cannot be accounted for because of
weak signals.  Therefore a `hybrid' method is employed.  Similar to
the method used in Refs \cite{compass_pgf,phd-robinet}, this approach
uses fits to the mass spectra which are sampled in bins of two
variables, NN signal purity (from a parameterisation without $a_{\rm
LL}$ in the training) and $fP_{\mu}a_{\rm LL}$.  The former sorts the
events according to their similar kinematic dependences, while the
latter is used to ensure the anticorrelation between $a_{\rm LL}$ and
the signal purity. The signal and background distributions belonging
to the mass spectra sampled in the bins are fitted by the same fitting
functions as defined above for describing the mass dependence of
signal and background.  Integrating the fits within the same mass
windows as used for the NN training procedure yields the signal
purities extracted from the fit.  For each of the two variables, a
function is built using linear interpolations between the fit results.
An iterative procedure is used to obtain a stable result on these two
functions simultaneously, and thereafter the correction $\lambda$ is
applied to the signal purity. Due to the statistical limitations, only
one parameterisation was built for each decay channel and year.
 
As the hybrid method can be used for both, low purity and high purity
channels, it was decided to use it for all parametrisations of signal
purities. Although more complex than the pure NN method, the hybrid
method results in a comparable statistical precision.

\subsection{Results on asymmetries}
\label{sec:asym_det}
The asymmetry $A^{\gamma {\rm N}}$ is extracted simultaneously with
the background asymmetry $A_{\rm B}^{\gamma {\rm N}}$ for each bin,
channel, and year of data taking except for low purity channels where
some data taking years of the same target set-up are merged.  Final
results sorted by D$^0$ decay mode are shown in
Tables~\ref{asym_results_DSD0subK} -- \ref{asym_results_DSK3pi}, where
$A^{\gamma {\rm N}}$ is given in each $(p_{\rm T}^{{\rm D}^0},E_{{\rm
D}^0})$ bin together with average values of kinematic variables. All
averages are calculated with the weight $w_{\rm S}^2$.  The
muon-nucleon asymmetry $A^{\mu {\rm N}\rightarrow \mu' {\rm D}^0
\rm{X} }$ can be obtained from $A^{\gamma {\rm N}}$ by multiplying it
by $\langle D\rangle$, which is a function of $\langle y\rangle$ and
$\langle Q^2\rangle$ and is also given in
Tables~\ref{asym_results_DSD0subK} -- \ref{asym_results_DSK3pi}.

As the extraction of $A^{\gamma {\rm N}}$ is performed based on event
weights, uncertainties introduced into the determination of the
asymmetry by each contribution to the weight have to be accounted
for. The uncertainty of $A^{\gamma {\rm N}}$ is acquired from a spread
of weighting factors $w$, Eq. (\ref{weight_ws}), which is obtained by
comparing the default analysis with weight $w_{0}$ to other analyses
with different weight, $w$.  The expectation value of the weighted
asymmetry is $\langle A^{\gamma {\rm N}}_{w_0} \rangle = \langle w w_0
\rangle / \langle w^2_0 \rangle A^{\gamma {\rm N}}$, as shown in
Ref.~\cite{celso_phd}.  The spread of $\langle w w_{0} \rangle /
\langle w_{0}^2 \rangle$ gives the relative systematic uncertainty of
$A^{\gamma {\rm N}}$.

The major sources of systematic uncertainties in the measurement of
$A^{\gamma {\rm N}}$ are discussed below.  The contributions from
$P_{\mu}$, $P_{\rm t}$ and $f$ are taken conservatively as 5\%, 5\%
and 2\% respectively, for both deuteron and proton targets.  In order
to study the contribution of $s/(s+b)$ to the systematic uncertainty,
three tests are performed. In one of them, different fitting functions
are used for the functional form of the background.  In the other two,
different mass windows are investigated concerning the Neural Network
parameterisation and the choice of the binning used in the
reconstruction of the D$^{0}$ spectra.  Note that each of these tests
leads to several new values of $s/(s+b)$ and, consequently, new
weights $w$ are obtained.  The resulting spread of weights is computed
for each year of data taking, each sample and each bin with respect to
the default weight $w_{0}$.  Thereafter, the weighted average of all
spreads is determined separately for each of the three systematic
tests considered.  The combined uncertainty on $s/(s+b)$ is obtained
from a quadratic superposition of these three uncertainties. The
resulting average value over all bins is 7\% of the measured
asymmetries.

The contribution of $D$ to the uncertainty of $A^{\gamma {\rm N}}$ is
obtained as follows.  According to the experimental uncertainty of 1\%
in the measurement of the mean value of the scattered muon momentum,
shifted values of $y$ are calculated for every event.  Thereafter, new
values of $D$ are computed from Eq. (\ref{depolarisation_fac}).  The
resulting spread of $\langle w w_{0} \rangle / \langle w_{0}^2
\rangle$ gives a systematic uncertainty of 1.6\%.

Systematic uncertainties of $A^{\gamma {\rm N}}$ arising from false
asymmetries and from the assumptions specified in
Eq. (\ref{reduction}) can be best estimated using the statistically
optimised method (see Sec. \ref{sec:results_new}).  First, they are
determined for the gluon polarisation $\Delta g/g$, and then they are
translated to $A^{\gamma {\rm N}}$ in bins of ($p_{\rm T}^{\rm D^{0}},
E_{\rm D^{0}}$) employing $\langle a_{\rm LL}/D \rangle$.  Averaging
over all bins, the resulting absolute values of the uncertainties due
to false asymmetries and the assumptions in Eq. (\ref{reduction}) are
0.022 and 0.007, respectively.

The total systematic uncertainties of $A^{\gamma {\rm N}}$, as given
in Tables~\ref{asym_results_DSD0subK}~--~\ref{asym_results_DSK3pi},
are obtained by adding all contributions in quadrature.

\begin{sidewaystable}[!h]
\begin{center}
\caption{Combined asymmetries $A^{\gamma {\rm N} \rightarrow {\rm D}^0
{\rm X}}$ for the D$^0_{{\rm K}\pi}$, D$^*_{{\rm K}\pi}$ and
D$^*_{{\rm K}_{{\rm sub}}\pi}$ samples in bins of ($p_{\rm T}^{{\rm
D}^0}$, $E_{{\rm D}^0}$), together with the weighted (with $w_{\rm
S}^2$) averages of several kinematic variables. First uncertainty is
statistical, second is systematic. \label{asym_results_DSD0subK}}
\setlength{\extrarowheight}{4pt}
\begin{tabular}{|c|c|c|c|c|c|c|c|}
\hline
\multicolumn{2}{|c|}{Bin limits} &  \multirow{2}{*}{$A^{\gamma {\rm N} 
\rightarrow {\rm D}^0 {\rm X}}$}  &    \multirow{2}{*}{$\langle y\rangle $}  &
$\langle Q^2\rangle $  & $\langle p_{\rm T}^{{\rm D}^0}\rangle $  &
$\langle E_{{\rm D}^0}\rangle $  & \multirow{2}{*}{$\langle D\rangle $}
\\ \cline{1-1} \cline{2-2}
 $p_{\rm T}^{{\rm D}^0}$ (GeV/$c$) &  $E_{{\rm D}^0}$ (GeV)         &
         &           &   (GeV/$c$)$^2$&
(GeV/$c$)       &   (GeV)         &    \\
\hline \hline
 0--0.3    & 0--30    &$-0.90 \pm 0.63 \pm 0.11$ &  0.50  & 0.46  & 0.19  & 24.3 & 0.62   \\
 0--0.3    & 30--50   &$-0.19 \pm 0.48 \pm 0.06$ &  0.60  & 0.69  & 0.20  & 39.1 & 0.74    \\
 0--0.3    & $> 50$   &$+0.07 \pm 0.68 \pm 0.06$ &  0.69  & 1.17  & 0.20  & 59.2 & 0.84    \\
 0.3--0.7  & 0--30    &$-0.18 \pm 0.37 \pm 0.04$ &  0.51  & 0.47  & 0.51  & 24.6 & 0.63    \\
 0.3--0.7  & 30--50   &$+0.10 \pm 0.26 \pm 0.04$ &  0.60  & 0.62  & 0.51  & 39.5 & 0.75    \\
 0.3--0.7  & $> 50$   &$-0.04 \pm 0.36 \pm 0.05$ &  0.69  & 0.73  & 0.51  & 59.0 & 0.83    \\
 0.7--1    & 0--30    &$-0.42 \pm 0.44 \pm 0.05$ &  0.50  & 0.45  & 0.85  & 24.7 & 0.62    \\
 0.7--1    & 30--50   &$-0.36 \pm 0.29 \pm 0.04$ &  0.61  & 0.60  & 0.85  & 39.2 & 0.75    \\
 0.7--1    & $> 50$   &$+1.49 \pm 0.42 \pm 0.15$ &  0.69  & 0.76  & 0.84  & 58.6 & 0.83    \\
 1--1.5    & 0--30    &$-0.30 \pm 0.35 \pm 0.03$ &  0.54  & 0.41  & 1.23  & 25.3 & 0.66    \\
 1--1.5    & 30--50   &$+0.13 \pm 0.23 \pm 0.01$ &  0.64  & 0.55  & 1.24  & 39.2 & 0.77    \\
 1--1.5    & $> 50$   &$-0.20 \pm 0.33 \pm 0.02$ &  0.71  & 0.73  & 1.24  & 58.3 & 0.85    \\
 $> 1.5$   & 0--30    &$+0.38 \pm 0.49 \pm 0.04$ &  0.56  & 0.47  & 1.84  & 25.6 & 0.69    \\
 $> 1.5$   & 30--50   &$~~0.00 \pm 0.25 \pm 0.02$& 0.65  & 0.70  & 1.92  & 39.9 & 0.79    \\
 $> 1.5$   & $> 50$   &$+0.36  \pm 0.33 \pm 0.04$ & 0.69  & 0.60  & 1.95  & 59.9 & 0.86    \\

\hline
\end{tabular}
\end{center}
\end{sidewaystable}

\clearpage
\begin{sidewaystable}[!h]
\begin{center}
\caption{Asymmetries $A^{\gamma {\rm N} \rightarrow {\rm D}^0 {\rm X}}$
for the $D^*_{{\rm K}\pi\pi^0}$ sample in bins of
($p_{\rm T}^{{\rm D}^0}$, $E_{{\rm D}^0}$) together with the weighted
(with $w_{\rm S}^2$) averages of several kinematic variables.
First uncertainty is  statistical, second is systematic.\label{asym_results_DSpi0}}
\setlength{\extrarowheight}{4pt}
\begin{tabular}{|c|c|c|c|c|c|c|c|c|}
\hline
\multicolumn{2}{|c|}{Bin limits}        &  \multirow{2}{*}{$A^{\gamma {\rm N} \rightarrow {\rm D}^0 {\rm X}}$}  &
\multirow{2}{*}{$\langle y\rangle $}  &  $\langle Q^2\rangle $  &
$\langle p_{\rm T}^{{\rm D}^0}\rangle $  &   $\langle E_{{\rm D}^0}\rangle$ &
\multirow{2}{*}{$\langle D\rangle $}
\\ \cline{1-1} \cline{2-2} $p_{\rm T}^{{\rm D}^0}$ (GeV/$c$) &
$E_{{\rm D}^0}$ (GeV)         &         &           &
(GeV/$c$)$^2$&(GeV/$c$)       &   (GeV)         &    \\
\hline \hline
 0--0.3    & 0--30    &$-0.63 \pm 1.29 \pm 0.08$ & 0.52  & 0.75  & 0.19 & 24.4& 0.65   \\
 0--0.3    & 30--50   &$+0.27 \pm 1.17 \pm 0.06$ & 0.67  & 0.65  & 0.20 & 38.8& 0.81   \\
 0--0.3    & $> 50$   &$-2.55 \pm 2.00 \pm 0.27$ & 0.72  & 1.12  & 0.19 & 59.3& 0.86   \\
 0.3--0.7  & 0--30    &$-0.24 \pm 0.80 \pm 0.04$ & 0.53  & 0.51  & 0.52 & 24.3& 0.65   \\
 0.3--0.7  & 30--50   &$+0.49 \pm 0.69 \pm 0.06$ & 0.65  & 0.65  & 0.51 & 39.0& 0.79   \\
 0.3--0.7  & $> 50$   &$-1.28 \pm 1.03 \pm 0.14$ & 0.72  & 0.77  & 0.51 & 59.1& 0.86   \\
 0.7--1    & 0--30    &$+0.55 \pm 0.95 \pm 0.06$ & 0.53  & 0.41  & 0.84 & 24.6& 0.65   \\
 0.7--1    & 30--50   &$-0.53 \pm 0.76 \pm 0.06$ & 0.63  & 0.53  & 0.86 & 39.4& 0.77   \\
 0.7--1    & $> 50$   &$-0.17 \pm 1.00 \pm 0.03$ & 0.73  & 0.80  & 0.85 & 58.2& 0.88   \\
 1--1.5    & 0--30    &$+1.35 \pm 0.86 \pm 0.14$ & 0.54  & 0.38  & 1.24 & 25.4& 0.67   \\
 1--1.5    & 30--50   &$-0.11 \pm 0.51 \pm 0.01$ & 0.64  & 0.59  & 1.25 & 39.6& 0.78   \\
 1--1.5    & $> 50$   &$-0.05 \pm 0.78 \pm 0.01$ & 0.74  & 0.62  & 1.25 & 58.3& 0.88   \\
 $> 1.5$   & 0--30    &$-0.19 \pm 1.14 \pm 0.03$ & 0.56  & 0.52  & 1.80 & 25.7& 0.70   \\
 $> 1.5$   & 30--50   &$-0.23 \pm 0.51 \pm 0.03$ & 0.66  & 0.66  & 1.88 & 40.0& 0.80   \\
 $> 1.5$   & $> 50$   &$+0.26 \pm 0.90 \pm 0.04$ & 0.74  & 0.88  & 1.92 & 57.3& 0.88   \\
\hline
\end{tabular}
\end{center}
\end{sidewaystable}

\clearpage

\begin{sidewaystable}[!h]
\begin{center}
\caption{Asymmetries $A^{\gamma {\rm N} \rightarrow {\rm D}^0 {\rm X}}$
for the $D^*_{{\rm K}\pi\pi\pi}$ sample in bins
  of ($p_{\rm T}^{{\rm D}^0}$, $E_{{\rm D}^0}$) together
with the weighted (with $w_{\rm S}^2$) averages of several kinematic
variables. First uncertainty is statistical, second is systematic.\label{asym_results_DSK3pi}}
\setlength{\extrarowheight}{4pt}
\begin{tabular}{|c|c|c|c|c|c|c|c|c|}
\hline
\multicolumn{2}{|c|}{Bin limits}        &  \multirow{2}{*}{$A^{\gamma {\rm N} \rightarrow {\rm D}^0 {\rm X}}$}  &
\multirow{2}{*}{$\langle y\rangle $}  &  $\langle Q^2\rangle $  &
$\langle p_{\rm T}^{{\rm D}^0}\rangle $  &   $\langle E_{{\rm D}^0}\rangle $  &
\multirow{2}{*}{$\langle D\rangle $}  \\
\cline{1-1} \cline{2-2} $p_{\rm T}^{{\rm D}^0}$ (GeV/$c$) &
$E_{{\rm D}^0}$ (GeV)         &         &           &   (GeV/$c$)$^2$&
(GeV/$c$)       &   (GeV)         &    \\
\hline \hline
 0--0.3    & 0--30    &$+7.03 \pm 4.74 \pm 0.71$ & 0.46  & 0.38  & 0.22 & 27.7 & 0.58   \\
 0--0.3    & 30--50   &$-2.05 \pm 1.10 \pm 0.21$ & 0.60  & 0.72  & 0.20 & 40.6 & 0.74   \\
 0--0.3    & $> 50$   &$+0.17 \pm 1.83 \pm 0.05$ & 0.69  & 0.88  & 0.20 & 59.1 & 0.84   \\
 0.3--0.7  & 0--30    &$-0.59 \pm 1.74 \pm 0.06$ & 0.52  & 0.31  & 0.53 & 27.8 & 0.71   \\
 0.3--0.7  & 30--50   &$+1.00 \pm 0.54 \pm 0.11$ & 0.61  & 0.44  & 0.52 & 39.7 & 0.80   \\
 0.3--0.7  & $> 50$   &$-1.75 \pm 0.84 \pm 0.18$ & 0.68  & 0.70  & 0.51 & 60.2 & 0.84   \\
 0.7--1    & 0--30    &$+2.91 \pm 2.61 \pm 0.30$ & 0.45  & 0.26  & 0.84 & 27.7 & 0.61   \\
 0.7--1    & 30--50   &$+1.42 \pm 0.57 \pm 0.15$ & 0.64  & 0.57  & 0.85 & 40.9 & 0.81   \\
 0.7--1    & $> 50$   &$+1.69 \pm 0.81 \pm 0.17$ & 0.69  & 0.58  & 0.86 & 60.9 & 0.84   \\
 1--1.5    & 0--30    &$-1.89 \pm 2.64 \pm 0.19$ & 0.46  & 0.31  & 1.22 & 27.7 & 0.64   \\
 1--1.5    & 30--50   &$-0.45 \pm 0.51 \pm 0.05$ & 0.63  & 0.58  & 1.23 & 41.1 & 0.79   \\
 1--1.5    & $> 50$   &$+1.06 \pm 0.66 \pm 0.11$ & 0.71  & 0.77  & 1.24 & 61.8 & 0.86   \\
 $> 1.5$   & 0--30    &$+1.64 \pm 3.52 \pm 0.17$ & 0.46  & 0.40  & 1.84 & 28.1 & 0.72   \\
 $> 1.5$   & 30--50   &$+0.44 \pm 0.68 \pm 0.05$ & 0.65  & 0.75  & 1.95 & 42.2 & 0.78   \\
 $> 1.5$   & $> 50$   &$+0.08 \pm 0.63 \pm 0.02$ & 0.74  & 0.77  & 2.03 & 64.4 & 0.88   \\
\hline
\end{tabular}
\end{center}
\end{sidewaystable}

\section{Determination of the gluon polarisation}
\label{sec:results_tot}
In this Section we present the results of our measurement of the gluon
polarisation. The extraction of $\Delta g/g$ from $A^{\gamma {\rm N}}$
at LO QCD accuracy is discussed in Sec. \ref{sec:results}.  The LO
determination of the gluon polarisation by a statistically optimised
method is described in Sec. \ref{sec:results_new}.  The extraction of
$\Delta g/g$ from $A^{\gamma {\rm N}}$ at NLO accuracy is presented in
Sec. \ref{sec:nlo}.

This analysis neglects any contribution from `intrinsic charm', {\it
i.e.} nonperturbative charm quark or charmed hadron components of the
nucleon wave function. Such contributions, estimated to be $\lapproxeq
1\%$ \cite{intrinsic_charm,alwall}, are fundamentally different from
the perturbative splitting of a gluon into a $c\bar c$ pair; the
latter decreases strongly with $x_{\rm B}$.  In the EMC measurement of
the charm component in the nucleon structure function $F_2^{c\bar c}$
\cite{emccharm}, a possible intrinsic charm contribution of about 1\%
at $x_{\rm B} \sim$ 0.4 could not be excluded \cite{emccharm,harris}.
Up to now, the estimates of Refs \cite{intrinsic_charm,alwall} cannot
be experimentally verified due to the poor statistics of the EMC
measurement at large $x_{\rm B}$, too low values of $x_{\rm B}$ in the
HERA $F_2^{c\bar c}$ measurements \cite{charm_hera}, and kinematic
acceptance limited to the region $x_{\rm B} \lapproxeq$ 0.1 for open
charm production in COMPASS.

The contribution of resolved-photon interactions was estimated using
the RAPGAP generator \cite{rapgap} and found to be negligible in our
kinematic domain.

\subsection{Leading Order results from the asymmetries}
\label{sec:results}
The information on the gluon polarisation contained in $A^{\mu {\rm N}
} $ can be decomposed at LO accuracy as
\begin{equation}
A^{\mu {\rm N} } = D A^{\rm \gamma N} = a_{\rm LL} \frac{\Delta
g}{g}~,
\label{lo_asym}
\end{equation}
assuming photon-gluon fusion as the underlying partonic process. Here
$a_{\rm LL}$ is the analysing power (also called `partonic asymmetry')
of the $\mu g \rightarrow \mu^\prime c\bar{c}$ process.

The analysing power $a_{\rm LL}$ depends on partonic kinematics.  It
is not accessible experimentally on an event-by-event basis.  It is
obtained using the Monte Carlo generator AROMA \cite{aroma} in leading
order QCD approximation, {\it i.e.} with parton showers switched off.
The generated D$^{0}$ events are processed with GEANT \cite{geant} to
simulate the full response of the COMPASS spectrometer, and then are
reconstructed with the same analysis chain as used for real events.
In order to provide $a_{\rm LL}$ values for real data, a Neural
Network with the same architecture as described in
Sec. \ref{sec:Method} is used to parameterise the generated $a_{\rm
LL}$ in terms of measured kinematic variables $X$.  Here the input
layer contains the following observables: $X = \{Q^2$, $y$, $x$,
$p_{\rm T}^{\rm D^0}$, $E_{\rm D^0}$\}.  As a result, $a_{\rm LL}(X)$
is obtained for real data on an event-by-event basis.

Contrary to the parameterisation of $s/(s+b)$, the Neural Network
predicts values for $a_{\rm LL}$ based on event kinematics. For each
generated event, the network tunes the strength of each
variable-neuron and neuron-neuron connection.  The strengths are
obtained by minimising the squared deviation between the expected
output, {\it i.e.} the generated $a_{\rm LL}$, and the actual Neural
Network prediction for $a_{\rm LL}$ based on $X$.  This training
process stops when the deviation between generated and parameterised
$a_{\rm LL}$ reaches a stable minimum.  Six separate $a_{\rm LL}(X)$
parameterisations were built: for three D meson decay channels
(K$\pi$, K$\pi\pi^0$, K$\pi\pi\pi$), each for two experimental
configurations (2002--2004 and 2006--2007).  The correlation achieved
between the generated and the parametrised analysing powers is 77\%
for the D$^*_{{\rm K}\pi\pi^0}$ channel and 82\% for the remaining
channels.  The trained network is applied to real data.

Knowing that $A^{\mu {\rm N}} = DA^{\gamma {\rm N}}$, the
determination of the gluon polarisation from Eq.~(\ref{lo_asym}) is
straightforward.  The extraction of $\Delta g/g$ from $A^{\gamma {\rm
N}}$ in bins of ($p_{\rm T}^{{\rm D}^0}$, $E_{{\rm D}^0}$) is
performed using the values of $\langle a_{\rm LL}/D \rangle$ shown in
Table~\ref{asym_results_all}.

\begin{table}[!h]
\begin{center}
\caption{The average LO photon-gluon asymmetries, $\left\langle a_{\rm LL}/D
\right\rangle$, in bins of
($p_{\rm T}^{{\rm D}^0}$, $E_{{\rm D}^0}$) for each D$^{ 0}$ decay
mode studied in the analysis. The averages use $a_{\rm LL}^{\rm LO}/D$
from data events, obtained from the Neural Network parameterisation;
they are weighted with $w_{\rm S}^2$.
\label{asym_results_all}}
\setlength{\extrarowheight}{8pt}
\begin{tabular}{|c|c|c|c|c|}
\hline
 \multicolumn{2}{|c|}
{\textbf{Bin limits}} &
\multicolumn{3}{|c|}
{\textbf{Photon-gluon asymmetries}} \\
\hline
{ \textbf{$p_{\rm T}^{{\rm D}^0}$}} & { \textbf{$E_{{\rm D}^0}$}}&
{ \textbf{${\rm D}^0_{{\rm K}\pi}$}, \textbf{${\rm D}^*_{{\rm K}\pi}$} and
\textbf{${\rm D}^*_{{\rm K}_{{\rm sub}}\pi}$} samples} &
{\textbf{${\rm D}^*_{{\rm K}\pi\pi^0}$} sample} &
{\textbf{${\rm D}^*_{{\rm K}\pi\pi\pi}$} sample}
\\
(GeV/$c$) & (GeV) &{ \textbf combined} &&
\\
\hline \hline
 0--0.3    & 0--30     & 0.65     &  0.62   &  0.64     \\
 0--0.3    & 30--50    & 0.68     &  0.65   &  0.63     \\
 0--0.3    & $> 50$    & 0.76     &  0.74   &  0.74     \\
 0.3--0.7  & 0--30     & 0.46     &  0.42   &  0.38     \\
 0.3--0.7  & 30--50    & 0.50     &  0.46   &  0.41     \\
 0.3--0.7  & $> 50$    & 0.56     &  0.53   &  0.52     \\
 0.7--1    & 0--30     & 0.26     &  0.19   &  0.25     \\
 0.7--1    & 30--50    & 0.26     &  0.21   &  0.25     \\
 0.7--1    & $> 50$    & 0.29     &  0.26   &  0.30     \\
 1--1.5    & 0--30     & 0.00     & $-0.06$ &  0.02     \\
 1--1.5    & 30--50    & 0.01     & $-0.05$ &  0.04     \\
 1--1.5    & $> 50$    & 0.05     & $-0.02$ &  0.08     \\
 $> 1.5$   & 0--30     & $-0.23$  & $-0.29$ & $-0.26$   \\
 $> 1.5$   & $30-50$   & $-0.26$  & $-0.31$ & $-0.23$   \\
 $> 1.5$   & $> 50$    & $-0.27$  & $-0.31$ & $-0.22$   \\
\hline
\end{tabular}
\end{center}
\end{table}

The gluon polarisation in LO QCD, obtained from $A^{\gamma {\rm N}}$ amounts to
\begin{equation}\label{finalres}
\left\langle\frac{\Delta g}{g}\right\rangle =-0.10\pm 0.22~(\mbox{stat.})
\pm 0.09~(\mbox{syst.})
\end{equation}
\noindent
in the range of $0.06< x < 0.22$ with a weighted $\langle x \rangle
\approx 0.11$, and a scale $\langle\mu^2\rangle \approx
13~(\GeV/c)^2$.  The range of $x$ is determined by the r.m.s. value of
a Gaussian distribution in $\log_{10}x$.  Assuming that ${\Delta
g}/{g(x)}$ is approximately a linear function of $x$ in the range
covered by the present data, the above result corresponds to the gluon
polarisation $\Delta g/g$ at the value $\langle x \rangle$.

The sources of systematic uncertainties in the measurement of $\Delta
g/g$ are listed in Table~\ref{tab:D02}.  The contributions from
$P_{\mu}$, $P_{\rm t}$, $f$, $s/(s+b)$ and $D$ are the same as
discussed in Sec. \ref{sec:asym_det}.  Contributions from false
asymmetries and from the assumption given in Eq.~(\ref{reduction}),
the same as in the statistically optimised method, are discussed in
Sec.~\ref{sec:results_new}.  In order to estimate the influence of the
simulation parameters on the determination of $a_{\rm LL}$, Monte
Carlo samples with different parameter sets are generated and the
analysing power is recalculated.  In these parameter sets, the mass of
the charm quark is varied between 1.3~GeV/$c^2$ and 1.6~GeV/$c^2$, and
the parton distribution functions as well as the factorisation scale
is varied by a factor of eight.  From each of these systematic tests a
new value of $\langle a_{\rm LL}/D \rangle$ is obtained, and
thereafter $\Delta g/g$ is recalculated for each ($p_{\rm T}^{{\rm
D}^0}$, $E_{{\rm D}^0}$) bin by dividing $A^{\gamma {\rm N}}$ by
$\langle a_{\rm LL}/D \rangle$.  The systematic uncertainty in each
bin is determined from the average spread of $\Delta g/g$ compared to
the result of the default analysis. The value for the systematic
uncertainty of gluon polarisation is obtained as a weighted average of
the systematic uncertainty in each bin.  The relative uncertainty
introduced by $a_{\rm LL}$ alone is 15\%.

\begin{table}[th]
\begin{center}
\caption{Contributions to the systematic
uncertainty of $\langle \Delta g/g \rangle^{\rm LO}$ obtained from $A^{\rm \gamma N}$. 
 \label{tab:D02}}
\begin{tabular}{|lc||lc|}
\hline
Source & $\delta \left (\langle  \Delta g/g \rangle \right )$ &
Source & $\delta \left (\langle  \Delta g/g \rangle \right )$\\
\hline
\hline
Beam polarisation $P_{\mu}$   & 0.005&$s/(s+b)$   & 0.007 \\
Target polarisation $P_{\rm t}$   &  0.005& False asymmetry   & 0.080 \\
Dilution factor $f$ & 0.002 & $a_{\rm LL}$& 0.015\\
Assumption, Eq.~(\ref{reduction})&0.025& Depolarisation factor $D$ & 0.002 \\
\hline
\multicolumn{4}{|c|}{Total uncertainty~~~ 0.086} \\
\hline
\end{tabular}
\end{center}
\end{table}

The final systematic uncertainty of $\langle \Delta g/g \rangle$
is obtained as a quadratic sum of all contributions. 
\clearpage

\subsection{Statistically optimised determination 
of the gluon polarisation at LO}
\label{sec:results_new}
The data described in Sec. \ref{sec:data_selection} allow for the
determination of $\langle \Delta g/g \rangle$ in a different,
statistically optimised way.  Practically it means that the gluon
polarisation is obtained by replacing the factor $D$ with
$a_{\textrm{LL}}$ in the definition of $w_{\rm S}$: $w_{S} =
P_{\mu}fa_{\textrm{LL}}s/(s+b)$.  The use of this weight allows us to
reproduce the results on $\langle \Delta g/g \rangle$ obtained from
$A^{\gamma {\rm N}} (\langle p_{\rm T}^{{\rm D}^{0}} \rangle, \langle
E_{{\rm D}^{0}} \rangle ) $ with about 6\% gain in the statistical
precision. This gain is due to a wide range of $a_{\rm LL}$ values but
the observed (anti)correlation between the signal purity and the
parameterised $a_{\textrm{LL}}$ has to be accounted for in the
parameterisation of $s/(s+b)$. This fact is crucial to obtain an
unbiased result of $\langle \Delta g/g \rangle$ with this
statistically optimised method.

Values for $\langle \Delta g/g \rangle$ and the background asymmetry
$\langle A_{\rm B}^{\gamma {\rm N}} \rangle$ were obtained for each of
the 48 weeks of data taking and separately for each of the five event
samples.  The results shown in Table~\ref{tab:Results} are the
weighted means of those values.

\begin{table}[!h]
\begin{center}
\caption{Results for $\langle \Delta g/g \rangle$ and 
$\langle A_{\rm B}^{\gamma {\rm N}} \rangle$ for each data sample.
Errors are statistical.\label{tab:Results}}
\begin{tabular}{|c|c|c|c|c|c|}
\hline
 & D$^{*}_{{\rm K}\pi}$ & D$^{*}_{{\rm K}\pi\pi^{0}}$ & D$^{*}_{{\rm K}\pi\pi\pi}$ & D$^{*}_{{\rm K}_{\rm sub}\pi}$ & D$^{0}_{{\rm K}\pi}$ \\
\hline
\hline
$\langle \Delta g/g \rangle$ & $-0.192 \pm 0.305$ & $-0.414 \pm 0.575$   
& ~~$0.614 \pm 0.667$  & ~~$0.497 \pm 0.995$  & ~~$0.020 \pm 0.415 $ \\ \hline 
$\langle A_{\rm B}^{\gamma {\rm N}} \rangle$ & $+0.019 \pm 0.029$ & $+0.051 \pm 0.035$  & $+0.004 \pm 0.036$  & $+0.004 \pm 0.047$  & $-0.005 \pm 0.004 $\\ \hline  
\end{tabular}
\end{center}
\end{table}

The value of the gluon polarisation is obtained as the weighted mean
of the five results shown in Table ~\ref{tab:Results} and amounts to
\begin{equation}
\left\langle\frac{\Delta g}{g}\right\rangle =-0.06\pm 0.21~(\mbox{stat.})\pm 0.08~(\mbox{syst.})
\label{direct_result}
\end{equation}
\noindent
in the range of $0.06< x < 0.22$ with a weighted $\langle x \rangle
\approx 0.11$, and a scale $\langle\mu^2\rangle \approx
13~(\GeV/c)^2$.

The major contributions for the systematic uncertainty given in
Eq. (\ref{direct_result}) are presented in Table~\ref{tab:D01}. They
were estimated as follows. In addition to $P_{\mu}$, $P_{\rm t}$, $f$
and $s/(s+b)$, the uncertainty of $a_{\rm LL}$ was also determined
from the spread of weights $\langle w w_{0} \rangle / \langle w_{0}^2
\rangle$ (where $w_{0}$ stands for the default analysis). The use of
different sets of parameters as described in Sec. \ref{sec:results}
gives rise to a relative systematic uncertainty of 9\% of the $\langle
\Delta g/g \rangle$ value originating from $a_{\rm LL}$. The relative
systematic uncertainty introduced by $s/(s+b)$ is 7\%.

In order to study the influence of false asymmetries, the D$^{*}_{{\rm
K}\pi}$ sample was divided into two sub-samples using criteria related
to the experimental apparatus, {\it e.g.} the slow pion going to the
left or to the right side of the incoming muon. The resulting
asymmetries were found to be compatible within their statistical
accuracies, {\it i.e.} no false asymmetries were observed. An upper
limit of the contribution of time dependent acceptance effects to the
systematic uncertainty was derived from the dispersion of $\langle
\Delta g/g \rangle$ and $\langle A_{\rm B}^{\gamma {\rm N}} \rangle$
in the 48 weeks of data taking.  The study was performed using the
background asymmetry, profiting from the large statistics. Then the
obtained results were translated to $\langle \Delta g/g \rangle$ using
the method described in Ref. \cite{celso_phd}. An uncertainty of 0.024
was obtained assuming that possible detector instabilities are similar
for background and signal events. Notice that the same assumption was
used to reduce the number of unknowns in Eq. (\ref{sum_w}) from 12 to
9.  Therefore, a more conservative approach is taken: the double ratio
of acceptances for the signal, $ \alpha_{\rm S}^{u} \cdot \alpha_{\rm
S}^{d'} / \alpha_{\rm S}^{d} \cdot \alpha_{\rm S}^{u'}$, is assumed to
be uncorrelated with the corresponding one for the background events.
The combination of these two cases leads to an upper limit of 0.08 for
the possible contribution of false asymmetries. This contribution is
also used in Sections \ref{sec:asym_det} and \ref{sec:results}.
\begin{table}[tp]
\begin{center}
\caption{Contributions to the systematic uncertainty of 
$\langle \Delta g/g \rangle^{\rm LO}$ obtained in a statistically
optimised method. \label{tab:D01}}
\begin{tabular}{|lc||lc|}
\hline
Source & $\delta \left (  \langle {\Delta g}/{g} \rangle \right )$ & 
Source & $\delta \left ( \langle  {\Delta g}/{g} \rangle \right )$\\
\hline
\hline
Beam polarisation $P_{\mu}$   & 0.003 &$s/(s+b)$   & 0.004\\
Target polarisation $P_{\rm t}$   &  0.003& $a_{\rm LL}$   & 0.005\\
Dilution factor $f$ &  0.001 & False asymmetry   & 0.080  \\
Assumption, Eq.~(\ref{reduction})&0.025&&\\
\hline
\multicolumn{4}{|c|}{Total uncertainty~~~0.084} \\
\hline
\end{tabular}
\end{center}
\end{table}

An uncertainty originating from the assumption for $\Delta g/g$,
analogous to that given in Eq.~(\ref{reduction}) and valid for $\Delta
g/g$ constant in the measured interval, 0.06 $< x < $0.22, is
estimated as follows. First a pair of extremal values of $\Delta g/g$
in that interval is selected as those given by the COMPASS NLO QCD fit
with $\Delta g > 0$, see Sec. \ref{sec:nlo}.  This fit is chosen to
maximise a potential influence of the above assumption. Next, a
difference between these two $\Delta g/g$ values is used as a bias in
the system of equations from which $\Delta g/g$ and $A^{\rm \gamma
N}_{\rm B}$ are determined.  The bias is added to all $\langle\Delta
g/g\rangle_{w_{\rm B} \beta_{\rm S}}$ terms while the $\langle\Delta
g/g\rangle_{w_{\rm S} \beta_{\rm S}}$ ones are left unchanged.  The
$\Delta g/g$ interval resulting from the equations gives a relative
systematic uncertainty due to the assumption.  Possible variations of
$A^{\rm \gamma N}_{\rm B}$ are studied in a similar way using a
parameterisation of the inclusive asymmetry, $A_1^{\rm d}$,
\cite{compass_lowq}. The systematic uncertainty on $\Delta g/g$ is
taken as the largest difference between the result obtained in the
default analysis and results of those tests.

\begin{figure}[!h]
\begin{center}
\includegraphics[width=0.75\hsize,clip]{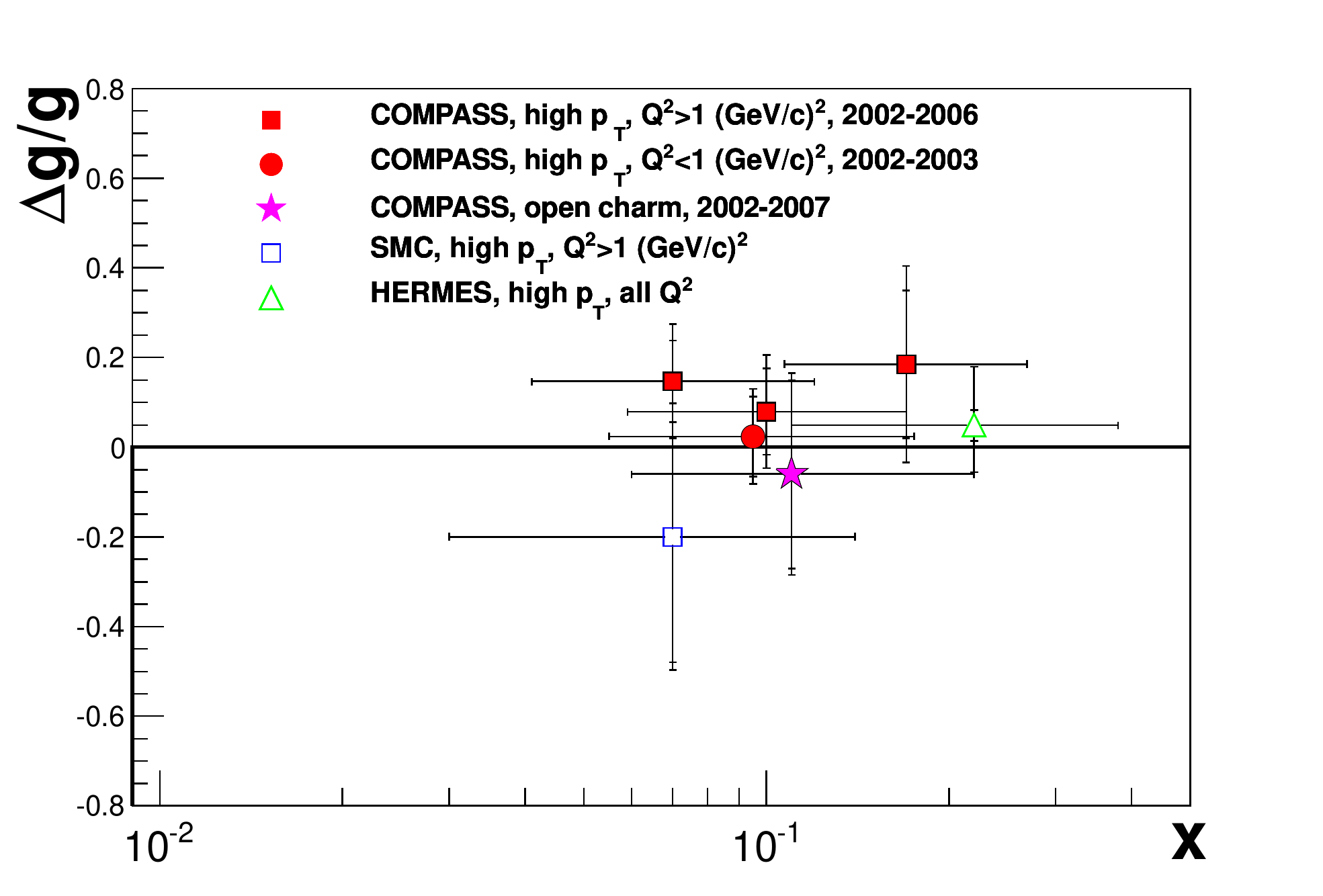}
\end{center}
\caption{ A compilation of gluon polarisation measurements from
open charm and high-$p_{\rm T}$ hadron production. The star denotes a
result of the present, open charm analysis, Eq. (\ref{direct_result}),
obtained at LO accuracy, 2002--2007 data and all values of $Q^2$.
Full squares denote a COMPASS result \cite{marcin} for high-$p_{\rm
T}$ hadron production on 2002--2006 data, for $Q^2 > $ 1 (GeV/$c$)$^2$
while a full circle corresponds to 2002--2003 data and $Q^2 < $ 1
(GeV/$c$)$^2$ \cite{compass_highpt_lowq}. The empty square shows the
SMC measurement \cite{SMC_highpt} for $Q^2 > $ 1 (GeV/$c$)$^2$ and the
empty triangle the HERMES result \cite{hermes_highpt} obtained for all
values of $Q^2$.  The horizontal bars mark the range in $x$ for each
measurement, the vertical ones give the statistical precision or the
total uncertainties.
} 
\label{fig:lo_results}
\end{figure}

The result given in Eq. (\ref{direct_result}) is shown in Fig.
\ref{fig:lo_results} together with a compilation of other LO gluon
polarisation measurements from high-$p_{\rm T}$ hadron production by
COMPASS \cite{marcin,compass_highpt_lowq}, SMC \cite{SMC_highpt} and
HERMES \cite{hermes_highpt}.  The present measurement is at a scale of
about 13 (GeV/$c$)$^2$ while other measurements are at 3
(GeV/$c$)$^2$.

\subsection{Next to Leading Order results}
\label{sec:nlo}
The extraction of the gluon polarisation as described in Secs
\ref{sec:results} - \ref{sec:results_new} was performed at LO where
the only process leading to open-charm production is PGF.  This
requires knowledge of the analysing power $a_{\rm LL}$ and the signal
strength on a bin-by-bin or event-by-event basis. Only combinatorial
background was considered in the LO analysis.

\begin{figure}[h]
\begin{center}
\includegraphics[width=0.5\hsize,clip]{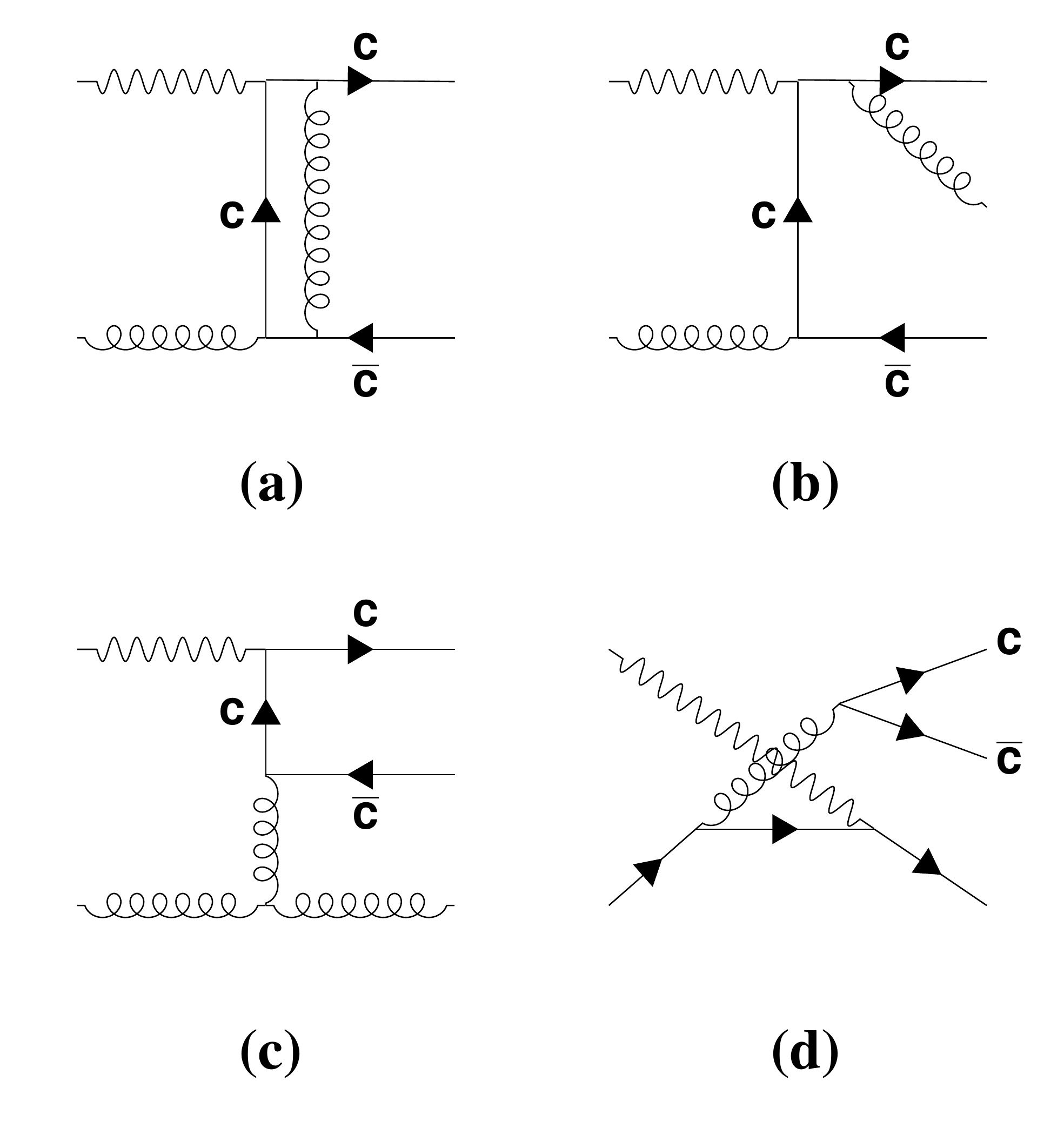}
\end{center}
\caption{ Examples of the NLO processes contributing to the
muoproduction of the $c\bar c$ pair:
a) virtual correction, b), c) gluon bremsstrahlung, d) light quark background
}
\label{fig:diagrams_nlo}
\end{figure}

In this section a brief outline of a method of computing the NLO QCD
corrections to the $a_{\rm LL}$ calculation in our analysis is given,
see also Ref. \cite{kk_nlo}.  Examples of the NLO processes
contributing to the muoproduction of the $c\bar c$ pair are shown in
Fig.\ref{fig:diagrams_nlo}.  Apart from the NLO corrections to the PGF
mechanism, Fig.\ref{fig:diagrams_nlo} a-c, there exists yet other NLO
contributions to muoproduction of open-charm, initiated by light
quarks; as an example a process where a gluon emitted by a light quark
creates the $c\bar c$ pair is shown in Fig. \ref{fig:diagrams_nlo}d.
Such processes do not probe the gluons inside the nucleon albeit they
contribute to the D meson signal. Therefore, in the extraction of the
gluon polarisation at NLO accuracy from the signal asymmetries, a
correction term has to be taken into account:
$$A^{\gamma{\rm N}} = \frac{a_{\rm LL}}{D} \frac{\Delta g}{g} + A_{\rm
corr}.$$ Here, $a_{\rm LL}$ is calculated at NLO accuracy and is
different from the corresponding one at LO.

The QCD calculations at NLO accuracy for spin averaged
\cite{beenakker} as well as polarisation dependent cross sections for
open charm production \cite{bojak} are available only at the
photoproduction limit, {\it i.e.} for $Q^2$ = 0. They are used in our
analysis to estimate the value of the NLO corrections to $a_{\rm LL}$
and the light quark contribution.  The average value of $Q^2$ in the
kinematic region of our measurement is about 0.6 (GeV/$c$)$^2$.  It
was confirmed by a direct check at LO accuracy that the $Q^2$ = 0
limit used in the calculation is a very good approximation in our
kinematic domain.

Note that only one D meson is registered in the COMPASS data, while
the second charm particle is unobserved. Also the NLO calculations of
Refs~\cite{beenakker,bojak} represent integrated cross sections for a
single charm quark (meson) observed in the final state. The cross
sections are integrated over the kinematic variables of the
`unobserved' second charm quark and radiated hard gluon, present in
NLO processes. The limits of the integration depend on the available
phase-space, left for `unobserved' partons which is determined by $x$
and the kinematics of the D-tagged $c$ quark.

In order to obtain $a_{\rm LL}$ in NLO accuracy on an event-by-event
basis, the AROMA generator (which is based on a LO matrix element)
with parton showers included is used, followed by a full simulation of
the detector.  In this way an approximation to the phase space needed
for NLO QCD corrections is provided. 

For every simulated event, the upper limit of the integration over the
energy of the unobserved gluon in the NLO emission process,
$\gamma^*g\rightarrow c\bar cg$, is obtained from the partonic
Mandelstam variables, $\hat s$ and $\hat t$.  Both variables are
calculated from the kinematics of simulated events. In particular
$\hat s$ is determined from $x$, $x_{\rm B}$ and $Q^2$ while $\hat t$
is related to the kinematics of the D-tagged charm quark.  The
integration over unobserved NLO real gluon emission reduces a
differential cross section for a three-body final state ($c\bar c g$)
to that for a two-body one ($c\bar c$), which has to be added to the
LO cross section ($c\bar c$, PGF) and the two-body virtual corrections
(see {\it e.g.}  diagram (a) in Fig.~\ref{fig:diagrams_nlo}).  In this way a
correct infra-red divergence cancellation is achieved \cite{bojak}.
The semi-inclusive partonic cross section at NLO accuracy is
calculated on an event-by-event basis for both spin averaged and spin
dependent case using formulae of Ref. \cite{bojak}, and consequently
$a_{\rm LL}$ at NLO accuracy is obtained.  The same procedure is
applied for the correction originating from a light quark. It should
be stressed that in this method of $a_{\rm LL}$ estimation at NLO,
only the values of $\hat s$ and $\hat t$ are taken from AROMA
simulated events.

To obtain the gluon polarisation at NLO accuracy, the measured
asymmetry for D meson production has to be combined with the $a_{\rm
LL}$ calculated at NLO. The kinematic variables $p_{\rm T}^{\rm D^0}$,
$E_{\rm D^0}$ and gluon momentum fraction $x$ define the total energy
of all particles produced in the final state of the partonic process,
including unobserved gluons emitted at NLO accuracy, $\gamma^* g
\rightarrow c\bar{c} g$.  However, in $a_{\rm LL}$ calculations,
simulated events at given $x$ values are used and the integration over
energy of the unobserved gluon is performed to obtain divergence-free
$a_{\rm LL}$, which depends on $\hat{s}$ and $\hat{t}$ only. Therefore
to be consistent with the NLO method of calculating $a_{\rm LL}$, the
asymmetry measured in ($p_{\rm T}^{\rm D^0}$, $E_{\rm D^0}$) intervals
is binned into five one-dimensional intervals of $p_{\rm T}^{\rm D^0}$
only.

\begin{table}[!h]
\begin{center}
\caption{The average values of the $\left\langle a_{\rm LL}/D
\right\rangle$ and $A_{\rm corr}$, at NLO and in bins of
$p_{\rm T}^{{\rm D}^0}$ for each D$^{ 0}$ decay
mode studied in the analysis. 
\label{tab:aLL_nlo}}
\setlength{\extrarowheight}{8pt}
\begin{tabular}{|c|c|c|c|c|c|c|}
\hline
{ \textbf{$p_{\rm T}^{{\rm D}^0}$}} & \multicolumn{2}{|c|} 
{ \textbf{${\rm D}^0_{{\rm K}\pi}$}, \textbf{${\rm D}^*_{{\rm K}\pi}$} and
\textbf{${\rm D}^*_{{\rm K}_{{\rm sub}}\pi}$} samples} &
\multicolumn{2}{|c|}{\textbf{${\rm D}^*_{{\rm K}\pi\pi^0}$} sample} &
\multicolumn{2}{|c|}{\textbf{${\rm D}^*_{{\rm K}\pi\pi\pi}$} sample}\\
(GeV/$c$) & \multicolumn{2}{|c|} {\textbf combined} &
\multicolumn{2}{|c|}{} & \multicolumn{2}{|c|}{}\\ 
\cline{2-7}
& $\langle a_{\rm LL}/D\rangle$& $A_{\rm corr}$
& $\langle a_{\rm LL}/D\rangle$& $A_{\rm corr}$
& $\langle a_{\rm LL}/D\rangle$& $A_{\rm corr}$\\
\hline \hline
 0.0--0.3  & $-0.130$ &  $0.001$  & $-0.127$  & $0.002$  & $-0.097$  & $0.000$ \\
 0.3--0.7  & $-0.241$ &  $0.003$  & $-0.263$  & $0.003$  & $-0.240$  & $0.001$ \\  
 0.7--1.0  & $-0.419$ &  $0.005$  & $-0.460$  & $0.004$  & $-0.404$  & $0.002$ \\
 1.0--1.5  & $-0.574$ &  $0.008$  & $-0.607$  & $0.008$  & $-0.572$  & $0.006$ \\
 $> 1.5$   & $-0.679$ &  $0.027$  & $-0.710$  & $0.020$  & $-0.719$  & $0.021$ \\
\hline
\end{tabular}
\end{center}
\end{table}

In each $p_{\rm T}^{\rm D^0}$ bin, the weighted averages
of $a_{\rm LL}/D$ and $A_{\rm corr}$ are calculated, Table \ref{tab:aLL_nlo}, 
and the gluon polarisation
is evaluated from the $A^{\gamma {\rm N}\rightarrow {\rm D}^0{\rm X}}$
asymmetries.
The NLO light quark contribution to the D meson asymmetry, $A_{\rm corr}$, 
is small, less than 5\%, compared to the measured asymmetries.
The gluon polarisation at NLO accuracy, obtained as weighted average
over all $p_{\rm T}^{\rm D^0}$ bins, is: 
\begin{equation}\label{finalres_nlo}
\left\langle\frac{\Delta g}{g}\right\rangle^{\rm NLO} =
 -0.13~ \pm~ 0.15~ ~(\mbox{stat.})  \pm 0.15~~(\mbox{syst.})
\end{equation}
\noindent
 It is determined in the interval 
$0.12 ~ < x < 0.33~$ with a weighted $\langle x \rangle \approx 0.20$,
at the scale $\langle \mu^2\rangle \approx$ 13 (GeV/$c$)$^2$. The gluon
momentum fraction $x$ is taken from the simulations.

\begin{figure}[h]
\begin{center}
\includegraphics[width=0.7\hsize,clip]{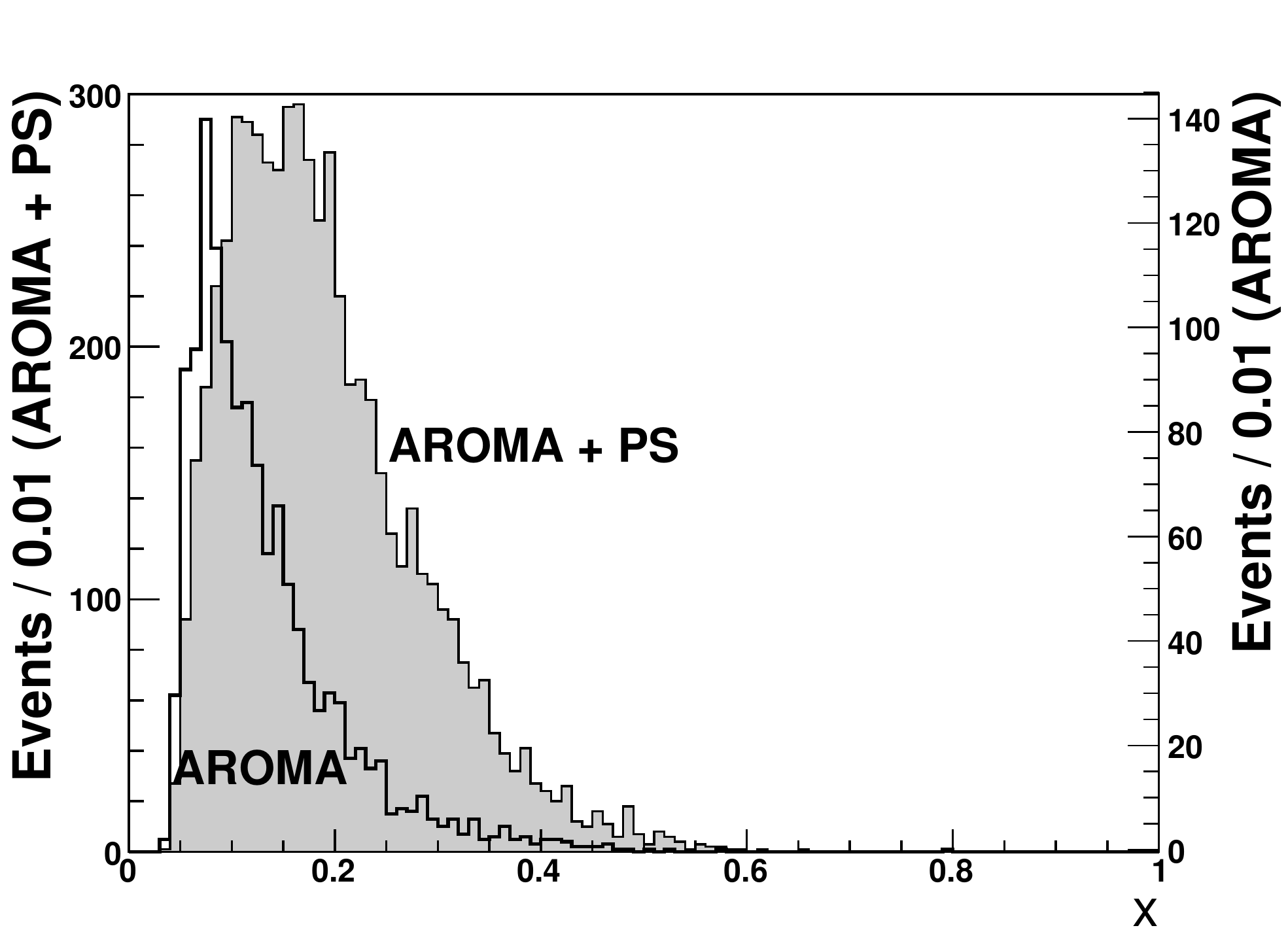}\\
\end{center}
\caption{ Distributions of the gluon momentum fractions $x$
for the simulated D$^*_{\rm K \pi}$ events at the LO accuracy
(marked `AROMA') and at LO with parton shower (AROMA + PS). 
Note different normalisations of the samples.
}
\label{fig:x_distr}
\end{figure}

For a given experimental acceptance for open charm tagging, the average
value of $x$ depends on the order of the QCD calculations used in the
analysis.  At the same time
the energy in the photon-gluon centre-of-mass system required to produce a
D$^0$ meson is higher with parton shower simulation as compared to the case
of LO where parton shower is not simulated. Therefore
a value of $\langle x \rangle$ at which the gluon polarisation is
determined at NLO $\langle x \rangle^{\rm NLO} \approx 0.20$,
is higher than $\langle x \rangle^{\rm LO} \approx 0.11$,
see for example Fig. \ref{fig:x_distr}.

A systematic uncertainty of the result in Eq. (\ref{finalres_nlo}) 
is estimated as follows, see Table \ref{tab:D03}.
Contributions from $P_\mu$, $P_{\rm t}$, $f$, $s/(s+b)$
and $D$ are the same as discussed in Sec. \ref{sec:asym_det}, 
{\it i.e.} 5\%, 5\%, 2\%, 7\% and 1.6\%
respectively. 
Contributions from false asymmetries and the assumption given in
Eq.~(\ref{reduction}),  
 the same as in the statistically optimised method at LO,
are discussed in Sec.\ref{sec:results_new}. 
Concerning the partonic asymmetry $a_{\rm LL}$, the following
contributions resulting from the NLO determination of $a_{\rm LL}$
were studied: its 
dependence on scale, the charm mass value and the Monte Carlo
mode (with or without PS). The renormalisation scale 
 (here chosen equal to the
factorisation one, Ref. \cite{bojak}) varied from $m_{\rm c}$ to
3$m_{\rm c}$. This changes the gluon polarisation at most by a factor of two,
leading to a conservative contribution to the systematic uncertainty of 0.1.
The variation of the charm quark mass between 1.3 and 1.6 GeV/$c^2$
results in a contribution of 0.05.
The systematic uncertainty contribution from the simulation method was
estimated using AROMA with and without parton showers. Note, that measured
D meson spectra are described reasonably well by AROMA both with and without
parton showers, and the difference in $a_{\rm LL}$ between both cases
is mainly due to the different phase space available for `unobserved'
partons. Such comparison allows us to give a conservative estimate of
uncertainty due to the simulation method equal to 0.04.
The total systematic uncertainty of $\langle \Delta g/g\rangle^{\rm NLO}$
is obtained by adding all the
 contributions in quadrature and amounts to 0.15.

\begin{table}[tp]
\begin{center}
\caption{Contributions to the systematic uncertainty of
$\langle \Delta g/g \rangle^{\rm NLO}$. \label{tab:D03}}
\begin{tabular}{|lc||lc|}
\hline
Source & $\delta \left ( \langle {\Delta g}/{g} \rangle \right )$ &
Source & $\delta \left ( \langle {\Delta g}/{g} \rangle \right )$\\
\hline
\hline
Beam polarisation $P_{\mu}$   & 0.006 &$s/(s+b)$   & 0.009\\
Target polarisation $P_{\rm t}$   &  0.006& $a_{\rm LL}$   & 0.119\\
Dilution factor $f$ &  0.003 & False asymmetry   & 0.080  \\
Assumption, Eq.~(\ref{reduction})&0.025&Depolarisation factor $D$&0.002\\
\hline
\multicolumn{4}{|c|}{Total uncertainty~~~0.146} \\
\hline
\end{tabular}
\end{center}
\end{table}

\begin{figure}[h!]
\begin{center}
\includegraphics[width=0.6\textwidth]{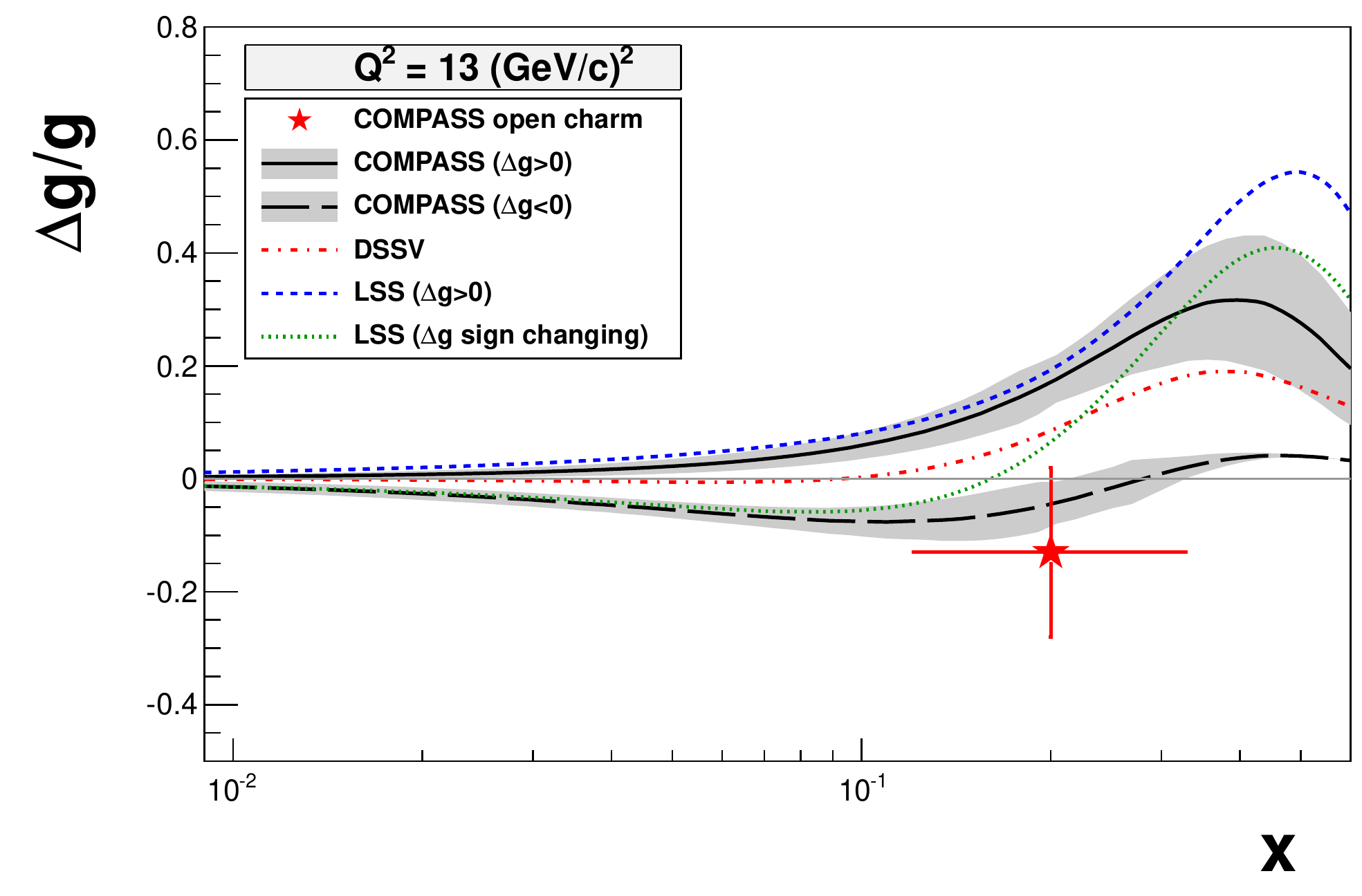}
\end{center}
\caption{ The present NLO measurement of the gluon polarisation
$\left <\Delta g(x)/g(x) \right >$ at $\langle\mu^2\rangle = $ 13
(GeV/$c$)$^2$, compared to the NLO QCD fits of COMPASS with $\Delta
g(x) > 0$ (continuous line) and $\Delta g(x) < 0$ (long-dashed) with
their respective error bands, of LSS \cite{lss_2010} (dashed and
dotted curves, respectively with $\Delta g > 0$ and $\Delta g$
changing sign) and of DSSV \cite{dssv_2009} (dashed-dotted curve), all
at the same value of $Q^2$ = 13 (GeV/$c$)$^2$.  The measurement error
and the error bands are statistical only; the horizontal bar marks the
range of $x$ in which $\left <\Delta g(x)/g(x) \right >$ is
determined.  }
\label{fig:dgg_nlo}
\end{figure}

The result on $\langle\Delta g/g\rangle^{\rm NLO}$,
Eq. (\ref{finalres_nlo}), was included in NLO QCD fits of polarised
parton distributions, see Appendix for details.  The fitted
distributions of $\Delta g(x)/g(x)$, evolved to $Q^2 = 13$
(GeV/$c$)$^2$, are shown in Fig. \ref{fig:dgg_nlo} together with error
bands corresponding to the statistical errors as derived from the
error matrix of the fitted parameters. The present NLO open charm
result agrees within 0.5 standard deviation with the fitted COMPASS
curve for $\Delta g(x) < 0$ and within 2 $\sigma$ with the one for
$\Delta g(x) > 0$.  It significantly influences the $\Delta g(x) > 0$
fit, reducing the value of $\Delta G$ from $0.39 \pm 0.07$ (stat.)  to
$0.22 \pm 0.08$ (stat.) at $Q^2$ = 3 (GeV/$c$)$^2$.

The results of two other global fits, DSSV \cite{dssv_2009} and LSS
\cite{lss_2010}, which employ both DIS and SIDIS asymmetries are also
shown in the same Figure.  In the DSSV fit, $\Delta g(x)$ changes sign
at $x \approx 0.1$ and is about 1.5 $\sigma$ above the COMPASS open
charm value.  In the case of LSS, two solutions, with positive and
with sign-changing $\Delta g(x)$ are quoted. The LSS fit cannot
distinguish between the positive and sign-changing $\Delta g(x)$
functions.  Both solutions give a positive $\Delta g(x)$ at the
$(x,Q^2)$ of the present measurement, about 2 and 1.3 $\sigma$ above
the measured value.

\section{Conclusions}
\label{sec:conclusions}
We have presented new results on the gluon polarisation in the nucleon
\DG ~and the virtual photon-nucleon asymmetries $A^{\gamma{\rm N}}$
obtained from charm production tagged by D meson decays in 160~GeV/$c$
polarised muon scattering off polarised proton and deuteron
targets. The results are based on a data sample collected between 2002
and 2007 and supersede the previously published ones
\cite{compass_pgf} as they are based on an enlarged data sample and an
improved analysis method.  The gluon polarisation is determined with
open charm production dominated by the photon-gluon fusion mechanism
followed by a spin-independent charm quark fragmentation into D
mesons. This analysis neglects any contributions from the intrinsic
charm; a contribution of resolved photon interactions was found
negligible.

Only one charmed meson is required in every event.  This meson is
selected through its decay in one of the following channels: D$^*$
(2010)$^+ \rightarrow$ D$^0\pi^+_{\rm slow}\rightarrow$
(K$\pi$/K$\pi\pi^0$/K$\pi\pi\pi$)$\pi^+_{\rm slow}$ or
D$^0\rightarrow$K$\pi$ .  The decays are selected using the invariant
mass distributions of identified kaons and pions.  A neural network is
used to distinguish signal from background events in the data. The
asymmetries $A^{\gamma {\rm N}}$ are extracted from these open charm
events in bins of D$^0$ transverse momentum and laboratory energy.
The average gluon polarisation obtained from these asymmetries at LO
QCD accuracy amounts to $\left\langle{\Delta g}/{g}\right\rangle^{\rm
LO} =-0.10\pm 0.22~(\mbox{stat.})\pm 0.09~(\mbox{syst.}).$ This result
is confirmed and statistically improved by employing a statistically
optimised method of extracting the gluon polarisation:
$$\left\langle\frac{\Delta g}{g}\right\rangle^{\rm LO} =-0.06\pm
0.21~(\mbox{stat.})\pm 0.08~(\mbox{syst.}).
$$ Both results are obtained in the range $0.06< x < 0.22$ of the
gluon momentum fraction with $\langle x \rangle \approx 0.11$ and a
scale $\langle\mu^2\rangle \approx 13~(\GeV/c)^2$.

For the first time in this analysis, next-to-leading order QCD
calculations for the determination of the gluon polarisation are
employed.  sing asymmetries $A^{\gamma{\rm N}}$ the gluon polarisation
is obtained as:
$$\left\langle\frac{\Delta g}{g}\right\rangle^{\rm NLO} = -0.13 \pm
0.15~(\mbox{stat.})  \pm 0.15~(\mbox{syst.}).
$$ In this case the range of $x$, $0.12 < x < 0.33$, leads to a higher
average value, $\langle x \rangle \approx 0.20$, while the scale is
approximately the same, $\langle \mu^2\rangle \approx$ 13
(GeV/$c$)$^2$.

The present measurement at LO QCD accuracy of the gluon polarisation
in the nucleon, together with other measurements of SMC, COMPASS and
HERMES, all situated around $x\sim 0.1$, points towards a small gluon
polarisation at that value of $x$. This is a hint for a small value of
the first moment, $\Delta G$, of the gluon helicity distribution,
although it in principle does not exclude a large value of $\Delta G$.

The $\left\langle{\Delta g}/{g}\right\rangle^{\rm NLO}$ result was
included in NLO QCD fits of polarised parton distributions.  It
significantly influences a fit in which $\Delta g(x) > 0$ was assumed,
reducing the value of $\Delta G$ from $0.39 \pm 0.07$ (stat.) to $0.24
\pm 0.09$ (stat.)  at $Q^2$ = 3 (GeV/$c$)$^2$, after it is included.

\section*{Appendix}
\label{sec:appendix}

In this appendix, new NLO QCD fits of polarised
parton distributions, including the
$\langle\Delta g/g\rangle^{\rm NLO}$ result of  Eq. (\ref{finalres_nlo})
are presented. In our previous fit \cite{compass} 
the gluon helicity distribution
is parametrised at a reference $Q^2$ of 3 (GeV/$c$)$^2$ in the form
\begin{equation}
 \Delta g(x) = \frac {\eta_{\rm g} \, x^{\alpha_{\rm g}}\, (1-x)^{\beta_{\rm g}}}
              {\displaystyle\int_0^1  x^{\alpha_{\rm g}}\, (1-x)^{\beta_{\rm g}}\, {\rm d}x}.
\label{parameterisation}
\end{equation}
Here $\eta_{\rm g}$ is the integral of $\Delta g(x)$,
$\eta_{\rm g}\equiv \Delta G$.
The same parameterisations,
Eq. (\ref{parameterisation}), are used for the singlet,
non--singlet quark and gluon helicity distributions except
for the singlet quark  in the fit with $\eta_g > 0$
where a factor $(1 + \gamma x)$ is added in order to allow a change of sign.
The high $x$ parameter of the gluon helicity distribution is fixed to 
$\beta_g = 10$ in this $\eta_g > 0$ fit so that the total number 
of free parameters remained equal to 10.
Two fits, in NLO QCD approximation and using all inclusive data
with $Q^2 > 1$ (GeV/$c$)$^2$ are performed: one with $\eta_{\rm g} > 0$, 
the other with $\eta_{\rm g} < 0$. Both of them gave 
a comparable $\chi^2$ probability.

In the new fit all the data used in Ref. \cite{compass} are employed
with an addition of the 15 COMPASS values of $A_1^p$ published later
\cite{compass_a1p}. As in the previous fit only statistical errors
are considered. 
The reference $Q^2$ is kept at 3 (GeV/$c$)$^2$ and the same parameterisations,
Eq. (\ref{parameterisation}), are used.
The total number of free parameters is also equal to 10.

The new open charm result is not attached to a precise value of $x$
 and thus its contribution is taken into account by the average
\begin{equation}
 \langle R_{\rm g}\rangle =  \frac{1}{(0.33 - 0.12)}\int_{0.12}^{0.33} 
\left [\frac{\Delta g}{g} (x,Q^2=13)\right ] {\rm  d}x 
\label{R}
\end{equation}
which is re--evaluated during the fit for any modification
of one of the gluon or singlet
quark parameters. The obtained value of $\langle R_{\rm g}\rangle$
is compared to the open charm result $v_{\rm OC} = -0.13$
with the statistical error $\sigma_{\rm OC} = 0.15$ and the $\chi^2$ of the fit
is incremented by $ (\langle R_{\rm g}\rangle - 
v_{\rm OC})^2/\sigma_{\rm OC}^2$.

The unpolarised gluon distribution $g(x,Q^2)$ in 
Eq. (\ref{R}) is taken from the
MRST04 parameterisation \cite{mrst04}. It was also used in Ref. \cite{compass}.
In contrast to previous parameterisations of the same group, MRST04 predicts
a slower decrease of the gluon distribution at high $x$, $(1 - x)^\beta$
with $\beta \sim $ 3--4. For this reason the choice of $\beta_{\rm g} = 10$ for
the fit with $\eta_{\rm g}> 0$ in Ref. \cite{compass} leads to a strongly
peaked distribution of $\Delta g/g$
which in turn generates a dip in the fitted distribution of $g_1^{\rm d}(x)$
around $x= 0.25$ for low values
of $Q^2$ and leads in some cases to very asymmetric errors due to the limits
imposed by the positivity condition $|\Delta g(x)| \le g(x)$.
To avoid these unphysical features, $\beta_{\rm g}$ is now  fixed to
6 in the fit with $\eta_{\rm g} > 0$.

The  present open charm result
has practically no effect on the fit for $\Delta g(x) < 0$, where
$\eta_{\rm g} = -0.34 \pm 0.12$ with- and without that measurement, while
it reduces significantly the positive $\eta_{\rm g}$, from
$\eta_{\rm g} = 0.39 \pm 0.07$ (stat.) to $\eta_{\rm g} = 0.22 \pm 0.08$ 
(stat.) at $Q^2$ = 3 (GeV/$c$)$^2$, after it is included.  
Similarily the values of $\alpha_{\rm g}$ for the fits including the open charm
point are $\alpha_{\rm g} = 1.31 \pm 0.47$ (stat.) for $\Delta g(x) > 0$
and $\alpha_{\rm g} = 0.26 \pm 0.48$ (stat.) for $\Delta g(x) < 0$,
both at $Q^2$ = 3 (GeV/$c$)$^2$.

\section*{Acknowledgments}
We gratefully acknowledge discussions with W. Vogelsang on the NLO
analysis of our data. We would like to thank the CERN management and staff
for their support, as well as
the skills and efforts of the technicians of the collaborating
institutes.

\end{document}

%% file: Authors2012CharmDG.tex
%
%

\section*{The COMPASS Collaboration}
\label{app:collab}

\begin{flushleft}
C.~Adolph\Irefn{erlangen},
M.G.~Alekseev\Irefn{triest_i},
V.Yu.~Alexakhin\Irefn{dubna},
Yu.~Alexandrov\Irefn{moscowlpi}\Deceased,
G.D.~Alexeev\Irefn{dubna},
A.~Amoroso\Irefn{turin_u},
A.A.~Antonov\Irefn{dubna},
A.~Austregesilo\Irefnn{cern}{munichtu},
B.~Bade{\l}ek\Irefn{warsaw},
F.~Balestra\Irefn{turin_u}
% ...
\end{flushleft}

%
%

\begin{Authlist}
\item \Idef{bielefeld}{Universit\"at Bielefeld, Fakult\"at f\"ur Physik, 33501 Bielefeld, Germany\Arefs{f}}
\item \Idef{bochum}{Universit\"at Bochum, Institut f\"ur Experimentalphysik, 44780 Bochum, Germany\Arefs{f}}
\item \Idef{bonniskp}{Universit\"at Bonn, Helmholtz-Institut f\"ur  Strahlen- und Kernphysik, 53115 Bonn, Germany\Arefs{f}}
\item \Idef{bonnpi}{Universit\"at Bonn, Physikalisches Institut, 53115 Bonn, Germany\Arefs{f}}
\item \Idef{brno}{Institute of Scientific Instruments, AS CR, 61264 Brno, Czech Republic\Arefs{g}}
\item \Idef{calcutta}{Matrivani Institute of Experimental Research \& Education, Calcutta-700 030, India\Arefs{h}}
\item \Idef{dubna}{Joint Institute for Nuclear Research, 141980 Dubna, Moscow region, Russia\Arefs{i}}
\item \Idef{erlangen}{Universit\"at Erlangen--N\"urnberg, Physikalisches Institut, 91054 Erlangen, Germany\Arefs{f}}
\item \Idef{freiburg}{Universit\"at Freiburg, Physikalisches Institut, 79104 Freiburg, Germany\Arefs{f}}
\item \Idef{cern}{CERN, 1211 Geneva 23, Switzerland}
\item \Idef{liberec}{Technical University in Liberec, 46117 Liberec, Czech Republic\Arefs{g}}
\item \Idef{lisbon}{LIP, 1000-149 Lisbon, Portugal\Arefs{j}}
\item \Idef{mainz}{Universit\"at Mainz, Institut f\"ur Kernphysik, 55099 Mainz, Germany\Arefs{f}}
\item \Idef{miyazaki}{University of Miyazaki, Miyazaki 889-2192, Japan\Arefs{k}}
\item \Idef{moscowlpi}{Lebedev Physical Institute, 119991 Moscow, Russia}
\item \Idef{munichlmu}{Ludwig-Maximilians-Universit\"at M\"unchen, Department f\"ur Physik, 80799 Munich, Germany\Arefs{f}\Arefs{l}}
\item \Idef{munichtu}{Technische Universit\"at M\"unchen, Physik Department, 85748 Garching, Germany\Arefs{f}\Arefs{l}}
\item \Idef{nagoya}{Nagoya University, 464 Nagoya, Japan\Arefs{k}}
\item \Idef{praguecu}{Charles University in Prague, Faculty of Mathematics and Physics, 18000 Prague, Czech Republic\Arefs{g}}
\item \Idef{praguectu}{Czech Technical University in Prague, 16636 Prague, Czech Republic\Arefs{g}}
\item \Idef{protvino}{State Research Center of the Russian Federation, Institute for High Energy Physics, 142281 Protvino, Russia}
\item \Idef{saclay}{CEA IRFU/SPhN Saclay, 91191 Gif-sur-Yvette, France}
\item \Idef{telaviv}{Tel Aviv University, School of Physics and Astronomy, 69978 Tel Aviv, Israel\Arefs{m}}
\item \Idef{triest_i}{Trieste Section of INFN, 34127 Trieste, Italy}
\item \Idef{triest}{University of Trieste, Department of Physics and Trieste Section of INFN, 34127 Trieste, Italy}
\item \Idef{triestictp}{Abdus Salam ICTP and Trieste Section of INFN, 34127 Trieste, Italy}
\item \Idef{turin_u}{University of Turin, Department of Physics and Torino Section of INFN, 10125 Turin, Italy}
\item \Idef{turin_i}{Torino Section of INFN, 10125 Turin, Italy}
\item \Idef{turin_p}{University of Eastern Piedmont, 15100 Alessandria,  and Torino Section of INFN, 10125 Turin, Italy}
\item \Idef{warsaw}{National Centre for Nuclear Research and University of Warsaw, 00-681 Warsaw, Poland\Arefs{n} }
\item \Idef{warsawtu}{Warsaw University of Technology, Institute of Radioelectronics, 00-665 Warsaw, Poland\Arefs{n} }
\item \Idef{yamagata}{Yamagata University, Yamagata, 992-8510 Japan\Arefs{k} }
\end{Authlist}
%
%
\vspace*{-\baselineskip}
\begin{Authlist}
\item \Adef{a}{Also at IST, Universidade T\'ecnica de Lisboa, Lisbon, Portugal}
\item \Adef{bb}{Supported by the DFG Research Training Group Programme 1102  ``Physics at Hadron Accelerators''}
\item \Adef{b}{Also at Chubu University, Kasugai, Aichi, 487-8501 Japan\Arefs{k}}
\item \Adef{c}{Also at KEK, 1-1 Oho, Tsukuba, Ibaraki, 305-0801 Japan}
\item \Adef{d}{On leave of absence from JINR Dubna}
\item \Adef{y}{present address: National Science Foundation, 4201 Wilson Boulevard, Arlington, VA 22230, United States}
\item \Adef{x}{present address: RWTH Aachen University, III. Physikalisches Institut, 52056 Aachen, Germany}
\item \Adef{e}{Also at GSI mbH, Planckstr.\ 1, D-64291 Darmstadt, Germany}
\item \Adef{f}{Supported by the German Bundesministerium f\"ur Bildung und Forschung}
\item \Adef{g}{Supported by Czech Republic MEYS Grants ME492 and LA242}
\item \Adef{h}{Supported by SAIL (CSR), Govt.\ of India}
\item \Adef{i}{Supported by CERN-RFBR Grants 08-02-91009}
\item \Adef{j}{\raggedright Supported by the Portuguese FCT - Funda\c{c}\~{a}o para a Ci\^{e}ncia e Tecnologia, COMPETE and QREN, Grants CERN/FP/109323/2009, CERN/FP/116376/2010 and CERN/FP/123600/2011}
\item \Adef{k}{Supported by the MEXT and the JSPS under the Grants No.18002006, No.20540299 and No.18540281; Daiko Foundation and Yamada Foundation}
\item \Adef{l}{Supported by the DFG cluster of excellence `Origin and Structure of the Universe' (www.universe-cluster.de)}
\item \Adef{m}{Supported by the Israel Science Foundation, founded by the Israel Academy of Sciences and Humanities}
\item \Adef{n}{Supported by the Polish NCN Grant DEC-2011/01/M/ST2/02350}
\item [{\makebox[2mm][l]{\textsuperscript{*}}}] Deceased
\end{Authlist}